%% file: paper.tex
\documentclass[acmsmall,screen,nonacm]{acmart}

\setcopyright{rightsretained}
\acmPrice{}
\acmDOI{10.1145/3591239}
\acmYear{2023}
\copyrightyear{2023}
\acmSubmissionID{pldi23main-p129-p}
\acmJournal{PACMPL}
\acmVolume{7}
\acmNumber{PLDI}
\acmArticle{125}
\acmMonth{6}
\received{2022-11-10}
\received[accepted]{2023-03-31}

\usepackage{booktabs}
\usepackage{amsmath}
\usepackage{subcaption}
\usepackage{bm}
\usepackage{mathpartir}
\usepackage{listings}
\usepackage[colorinlistoftodos]{todonotes}
\usepackage{paralist}

\usepackage{hyperref}

\usepackage{algorithm}
\usepackage{algpseudocode}
\algtext*{EndWhile}
\algtext*{EndIf}
\algtext*{EndFor}
\algtext*{EndProcedure}

\definecolor{mygreen}{rgb}{0,0.6,0}
\definecolor{mygray}{rgb}{0.5,0.5,0.5}
\definecolor{mymauve}{rgb}{0.58,0,0.82}

\lstdefinelanguage{Souffle}{
  sensitive,
  morecomment=[l]{//},
  morecomment=[s]{/*}{*/},
  moredelim=[s][{\itshape\color[rgb]{0,0,0.75}}]{\#[}{]},
  morestring=[b]{"},
  alsodigit={},
  alsoother={},
  alsoletter={!},
  otherkeywords={=>},
  otherkeywords={:-},
  otherkeywords={|},
  otherkeywords={.},
  morekeywords={decl, type}
}%
\lstdefinelanguage{egglog}{
  sensitive,
  morecomment=[l]{;},
  moredelim=[s][{\itshape\color[rgb]{0,0,0.75}}]{\#[}{]},
  morestring=[b]{"},
  alsodigit={},
  alsoother={},
  alsoletter={!},
  alsoletter={:},
  alsoletter={-},
  otherkeywords={=>},
  otherkeywords={:-},
  otherkeywords={|},
  otherkeywords={.},
  morekeywords={assert, function, relation, datatype, rewrite,
                rule, union, merge, default, when,
                check, set, run, sort, define, extract,
                old, new},
  otherkeywords={:merge, :default, :when}
}%

\newcommand{\rw}[1]{\todo[inline,color=yellow]{\textsf{#1} \hfill \textsc{--Remy}}}

\input{macros.tex}
\input{egglogdatasmall/macros.tex}

\input{egglogdatabignodelimit/macros.tex}
\begin{document}

\title{Better Together: Unifying Datalog and Equality Saturation}

\author{Yihong Zhang}
\orcid{0009-0006-5928-4396}
\affiliation{
  \institution{University of Washington}
  \country{USA}
}
\author{Yisu Remy Wang}
\orcid{0000-0002-6887-9395}
\affiliation{
  \institution{University of Washington}
  \country{USA}
}
\author{Oliver Flatt}
\orcid{0000-0002-0656-235X}
\affiliation{
  \institution{University of Washington}
  \country{USA}
}
\author{David Cao}
\orcid{0000-0002-6163-1821}
\affiliation{
  \institution{UC San Diego}
  \country{USA}
}
\author{Philip Zucker}
\orcid{0000-0003-0244-5962}
\affiliation{
  \institution{Draper Laboratory}
  \country{USA}
}
\author{Eli Rosenthal}
\orcid{0009-0008-9386-1614}
\affiliation{
  \institution{Google}
  \country{USA}
}
\author{Zachary Tatlock}
\orcid{0000-0002-4731-0124}
\affiliation{
  \institution{University of Washington}
  \country{USA}
}
\author{Max Willsey}
\orcid{0000-0001-8066-4218}
\affiliation{
  \institution{University of Washington}
  \country{USA}
}

\begin{abstract}
  \input{abstract}
\end{abstract}

\ccsdesc[300]{Theory of computation~Constraint and logic programming}
\ccsdesc[500]{Theory of computation~Equational logic and rewriting}

\keywords{Program optimization, Rewrite systems, Equality saturation, Datalog}

\maketitle

\renewcommand{\shortauthors}{Y.~Zhang, Y.~R.~Wang, O.~Flatt, D.~Cao, P.~Zucker, E.~Rosenthal, Z.~Tatlock, and M.~Willsey}

\input{intro.tex}

\input{background.tex}
\input{overview.tex}
\input{semantics.tex}

\input{implementation.tex}
\input{case-study.tex}
\input{relatedwork.tex}
\input{conclusion.tex}
\begin{acks}
We thank the anonymous reviewers for their thoughtful feedback.
We are grateful to Martin Bidlingmaier 
 for sharing his insights on \texttt{EqLog}, a concurrent work to \egglog,
 to Martin Bravenboer for discussions on
 the \texttt{Rel} programming language,
 to Scott Moore and Langston Barrett
 for answering questions about \cclyzerpp,
 and to friends at the UW PLSE group 
 for their feedback on the early draft.
  This material is based upon work supported by the National Science Foundation under Grant No. 1749570, by the U.S. Department of Energy under Award Number DE-SC0022081, by DARPA under contract FA8650-20-2-7008, and by the Applications Driving Architectures (ADA) Research Center, a JUMP Center co-sponsored by SRC and DARPA.
\end{acks}

\bibliography{references}

\appendix
\newpage
\input{pearls.tex}
\input{sn.tex}

\end{document}

%% file: macros.tex
\usepackage{xspace}

\newcommand{\Egraph}{\mbox{E-graph}\xspace}
\newcommand{\egraph}{\mbox{e-graph}\xspace}
\newcommand{\Egraphs}{\mbox{E-graphs}\xspace}
\newcommand{\egraphs}{\mbox{e-graphs}\xspace}

\newcommand{\eclass}{\mbox{e-class}\xspace}

\newcommand{\eclasses}{\mbox{e-classes}\xspace}
\newcommand{\ematch}{\mbox{e-match}\xspace}
\newcommand{\ematching}{\mbox{e-matching}\xspace}

\newcommand{\Enodes}{\mbox{E-nodes}\xspace}
\newcommand{\enodes}{\mbox{e-nodes}\xspace}
\newcommand{\enode}{\mbox{e-node}\xspace}

\newcommand{\egg}{\texttt{egg}\xspace}
\newcommand{\eqrel}{\texttt{eqrel}\xspace}
\newcommand{\souffle}{Souffl\'e\xspace}
\newcommand{\egglog}{\texttt{egglog}\xspace}
\newcommand{\Egglog}{\texttt{egglog}\xspace}
\newcommand{\cclyzerpp}{\texttt{cclyzer++}\xspace}
\newcommand{\eqsat}{\mbox{EqSat}\xspace}
\newcommand{\set}[1]{\{#1\}}                    
\newcommand{\N}{\mathbb{N}}

\newcommand{\seminaive}{semi-na\"ive\xspace}
\newcommand{\naive}{na\"ive\xspace}
\newcommand{\merge}{\lstinline[language=egglog]{:merge}\xspace}
\def\ngets{\textit{ :- }}

\newcommand{\revision}[1]{#1}
\newcommand{\revdel}[1]{}

%% file: egglogdatasmall/macros.tex
\newcommand{\numtests}{289\xspace}
\newcommand{\numbettererror}{104\xspace}
\newcommand{\numworseerror}{135\xspace}

\newcommand{\timeegglogminutes}{73.91\xspace}
\newcommand{\timevanillaminutes}{81.91\xspace}

%% file: abstract.tex
We present \egglog,
 a fixpoint reasoning
 system that unifies
 Datalog and equality saturation (\eqsat).
Like Datalog,
 it supports efficient incremental execution,
 cooperating analyses,
 and lattice-based reasoning.
Like \eqsat,
 it supports term rewriting,
 efficient congruence closure,
 and extraction of optimized terms.

We identify two recent applications---%
 a unification-based pointer analysis in Datalog
 and an \eqsat-based floating-point term rewriter---%
 that have been hampered
 by features missing from Datalog but found in \eqsat
 or vice-versa.
We evaluate \egglog by reimplementing
 those projects in \egglog.
The resulting systems in \egglog are 
 faster, simpler, and fix bugs 
 found in the original systems.


%% file: intro.tex
\section{Introduction}


Equality saturation (\eqsat) and Datalog are both
 fixpoint reasoning frameworks with 
 many 
 applications,
 extensions,
 and high-quality implementations~\cite{egg, souffle}.
They share a common setup: 
 the user provides \emph{rules} and an initial set of \emph{facts} 
   (a term in \eqsat and a database in Datalog),
 then the system derives a larger and larger set of facts from those inputs.
However,
 their commonalities have not---until now---been fully realized or exploited.
As a result, 
 the frameworks have developed independently and are used in different domains.
Datalog is well-studied by the databases community, 
 and practitioners use modern implementations
 to build program analyses%
 ~\cite{cclyzer,cclyzerpp,doop-datalog}.
Equality saturation
 is a more recent, term-centric technique
 favored in the programming languages community
 for program optimization and verification.

As users apply EqSat and Datalog to new, 
  more demanding problems,
  the limitations of each tool
  become apparent.
For example,
 Herbie~\cite{herbie}, 
 a tool that uses \eqsat to optimize floating-point accuracy, 
 relies on unsound rewrites because it lacks the analyses
 to prove that certain rewrites are safe 
 (e.g. $x/x \rightarrow 1$ only if $x \neq 0$).
To combat the unsoundness,
 Herbie must validate the results of \eqsat
 and discard them if unsoundness was detected.
On the Datalog side,
 \cclyzerpp~\cite{cclyzerpp},
 a recent points-to analysis system implemented in Datalog
 that supports Steensgaard analyses~\citep{DBLP:conf/popl/Steensgaard96}
 for LLVM~\citep{llvm}
 resorted to an ad-hoc implementation of union-find,
 because the provided implementation of equivalence relations
 was too slow.
The resulting implementation's complexity led to bugs
 in the pointer analysis.
In short, EqSat struggles to support rich analyses, 
  and equational reasoning in Datalog is complex and slow.

\emph{Our key insight is that
 the efficient equational reasoning of EqSat and 
 the rich, composable semantic analyses of Datalog 
  make up for each other's weaknesses, 
  and unifying the two paradigms brings 
  together---and goes beyond---the best of both worlds.}
In fact, spontaneous developments in both communities 
  have already begun converging towards each other:
Datalog tools have 
  added efficient equivalence relations~\cite{eqrel},
  lattices~\cite{flix,ascent},
  and some support for datatypes~\cite{souffle-adt}, 
  while the \eqsat community
  has recently developed support for
  conditional rewriting,
  lattice-based analyses~\cite{egg,metatheory.jl},
  and relational pattern matching~\cite{relational-ematching}.
We bring this trend to completion and close the gap
  between EqSat and Datalog.

In this work,
 we propose \egglog,
 a fixpoint reasoning system that subsumes both \eqsat and Datalog.
It contains all of the innovations listed above as well as new ones,
 and it addresses crucial limitations that have prevented progress in 
 real-world applications.
\egglog is essentially a Datalog engine
 with two main extensions. 
First,
 \egglog has a built-in, extensible notion of equality.
The user can assert that two terms are equivalent,
 from which point on they are indistinguishable to the system.
For example,
 consider a relation with a single tuple: $R = \set{(a, b)}$
 for distinct $a$ and $b$.
The query $R(x, x)$ would yield nothing,
 but if the user asserts that $a$ and $b$ are equivalent,
 then the query would return the equivalence class containing both $a$ and $b$.
Second,
 \egglog has built-in support for (uninterpreted) functions.
From a relational perspective,
 a function is a relation with a functional dependency 
 from its arguments to its output,
 i.e., the output is uniquely determined by the arguments.
However, user-extensible equality
 introduces challenges for maintaining functional dependencies.
Consider a function $f$ such that $f(a) = b$ and $f(c) = d$, 
 but $b$ does not (yet) equal $d$.
What happens when the user asserts that $a$ and $c$ are equivalent?
An \egglog function can be annotated with a \emph{merge expression},
 a novel mechanism that 
 \egglog uses to resolve functional dependency violations
 by combining the two conflicting output values.
In the above case,
 $f$'s merge expression might 
 assert that $b = d$ (essentially asserting congruence of $f$),
 or return the supremum of $b$ and $d$.
The flexibility of merge expressions allows \egglog
 to exceed the expressive power of both \eqsat and 
 Datalog extensions with lattices.
The high-level \Egglog language allows the user to specify 
  complex interactions among terms, equivalence classes, 
  and lattice values.
At the same time, highly optimized algorithms for 
  relational and equational reasoning 
  work together to make \Egglog efficient.

The combination of \eqsat and Datalog
 also brings many practical---if somewhat more prosaic---benefits.
For example,
 \citet{relational-ematching}
 observed that \eqsat is hampered by inefficient
 \ematching (pattern matching modulo equality) algorithms,
 and that a relational approach can be vastly more efficient.
\egglog's Datalog-first design naturally 
 supports efficient \ematching by reducing it to a relational query.
This goes even further:
incremental \ematching is only supported in some SMT solvers like Z3~\cite{z3}
 and has not yet made its way into \eqsat implementations,
 while \egglog supports them \emph{for free} with semi-na\"ive evaluation~\citep{seminaive},
 a common technique that makes Datalog incremental.
\egglog's support for functions
 provides the basis for working with terms,
 which only have limited support in other Datalog systems~\cite{souffle-adt}.
Users can also define multiple
 functions and datatypes
 to model their domain,
 unlike most \eqsat tools~\cite{egg,metatheory.jl}
 where users are forced to use a single, ad-hoc datatype.
Finally,
 \egglog is designed as a language (as well as a library),
 making it more accessible
 than \eqsat libraries~\cite{egg,metatheory.jl} 
 that are locked to their implementation language.

We perform two case studies showing that \egglog out-performs 
  state-of-the-art applications of \eqsat and Datalog respectively.
First, we show that \egglog makes Steensgaard-style points-to analyses
  faster and easier to write.
Compared to the \souffle Datalog system,\xspace \egglog computes the points-to analysis
  4.96$\times$ faster.
Second, we demonstrate the power of \egglog
 with a new, sound implementation of Herbie's \eqsat procedure.
This allows Herbie to perform aggressive optimizations soundly,
  and return results faster given the same error tolerance.


In summary, this paper makes the following contributions:
\begin{compactitem}
  \item We introduce a bottom-up, Datalog-like logic language 
   for equality saturation and similar unification-based algorithms.
  \item We present a fixpoint semantics for the core language of \egglog.
  \item We present an implementation for \egglog with optimizations 
   from database research such as semi-na\"ive evaluation.
\end{compactitem}

%% file: background.tex
\section{Background}

\Egglog is designed as a Datalog variant
 with extensions that make it subsume \eqsat.
This section will introduce both Datalog and \eqsat
 in their own terms,
 while \autoref{sec:overview}
 will show how they both fit within the \egglog framework.

\subsection{Datalog}
\label{sec:datalog}

\autoref{fig:datalog-tc}
 shows a Datalog program to compute the transitive closure of a graph.
Datalog is a recursive database query language that
 represents data as \emph{relations}.
Each relation is a set of tuples, and all tuples in 
  the same relation share the same arity.
A Datalog program consists of a set of \emph{rules}.
Each rule is a \emph{conjunctive query} of the form
\(
  Q(\mathbf{x}) \ngets 
  R_1(\mathbf{x}_1), R_2(\mathbf{x}_2), \ldots, R_n(\mathbf{x}_n)
\)
where each $\mathbf{x}$ and $\mathbf{x}_i$ is a tuple of variables or constants. 
The atom $Q(\mathbf{x})$ is called the \emph{head} of the rule, 
  and the atoms $R_i(\mathbf{x}_i)$ comprise the \emph{body}.
The body binds variables to be used in the head to create new facts;
 all variables in the head must appear in the body.
Specifically, running a rule adds the following facts:
\(
  \{
    Q(\mathbf{x}[\sigma]) \mid
    \bigwedge_i \mathbf{x}_i[\sigma] \in R_i
  \}
\),
where $\sigma$ is a substitution 
 that maps all the variables in the rule to constants.
In other words,
 querying the body creates substitutions such that
 every substituted body atom is in the database;
 these substitutions are then applied to the head to create new facts.

Each rule can be seen as a function 
 from the current database to a new database that includes
 the facts created by the rule;
 call this function $T_r$ for some rule $r$.
The set of all rules $r$ in a Datalog program $p$ 
 therefore defines a function $T_p$ 
 from the current database to a new database: 
 $T_p(\mathsf{DB}) = \bigcup_{r \in p} T_r(\mathsf{DB})$.
This function is called the immediate consequence operator (ICO) of the program,
  which we denote $T_p$.
To run a Datalog program, we start with an empty database and
 repeatedly apply $T_p$ until the database stops changing.
A fundamental result in Datalog is that every program terminates, 
  and the final result is the least fixpoint of $T_p$~\citep{alice-book}.


\begin{figure}
\begin{subfigure}[b]{0.4\linewidth}
\begin{align*}
E(1, 2)&. \\
E(2, 3)&. \\
E(3, 4)&. \\
TC(x, y) & \ngets E(x, y). \\
TC(x, y) & \ngets TC(x, z), E(z, y).
\end{align*}
\caption{
  Transitive closure in Datalog.
  Facts (e.g. $E(1, 2)$) are given as rules without bodies.
}\label{fig:datalog-tc-code}
\end{subfigure}
\hfill
\begin{subfigure}[b]{0.55\linewidth}
\begin{tabular}{lll}
    iter\# & E & TC \\
    \hline 
    0 & $\emptyset$ & $\emptyset$ \\
    1 & $\set{(1,2),(2,3),(3,4)}$ & $\emptyset$ \\
    2 & $\set{\ldots}$ & $\set{(1,2),(2,3),(3,4)}$ \\
    3 & $\set{\ldots}$ & $\set{\ldots, (1,3),(2,4)}$ \\
    4 & $\set{\ldots}$ & $\set{\ldots, (1,4)}$ \\
\end{tabular}
\caption{
  Execution trace of transitive closure. 
  ``$\ldots$'' includes tuples from the cell directly above.
}
\label{fig:datalog-tc-exec}
\end{subfigure}
\caption{
  Transitive closure is the classic Datalog example.
  It iteratively computes the transitive closure ($TC$) 
  of an edge relation $E$
  by applying the rules in \autoref{fig:datalog-tc-code}.
}\label{fig:datalog-tc}
\end{figure}


Datalog became popular in programming languages research 
  as a declarative language for specifying large-scale program analyses
  such as points-to analyses~\cite{doop-datalog}.
In order to support abstract interpretation-style analyses,
  researchers have extended Datalog to work over lattices.
In the lattice semantics,
 a relation is viewed as a function from tuples to a lattice,
We then generalize Datalog rules to be over functions:
\(
  Q(\mathbf{x}) \!\mapsto\! x
  \ngets 
  R_1(\mathbf{x_1}) \!\mapsto\! x_1, 
  \cdots, 
  R_n(\mathbf{x_n})\!\mapsto\! x_n
\).
The value of $Q$ on input $\mathbf{x}$ is 
 the supremum of valid $x$s producible by the body, i.e.,
$Q(\mathbf{x})=
  \bigsqcup
  \{
    x \mid
    \bigvee_{\mathbf{x}_\text{free}}
    R_1(\mathbf{x_1})=x_1 \land
    \ldots\land
    R_n(\mathbf{x_n})=x_n
  \}
$
where $\mathbf{x}_\text{free}$ is the set of variables in the body 
 that do not appear in the head and $\sqcup$ is 
 a lattice join (i.e., supremum) operator.
\Egglog's support for lattices
 is motivated by other 
 modern Datalog implementations~\cite{datalogo,ascent,flix}
 that support this extension.

\subsection{Equality Saturation}

\begin{figure}
    \begin{subfigure}[t]{0.30\linewidth}
        \centering
        \includegraphics[scale=0.5]{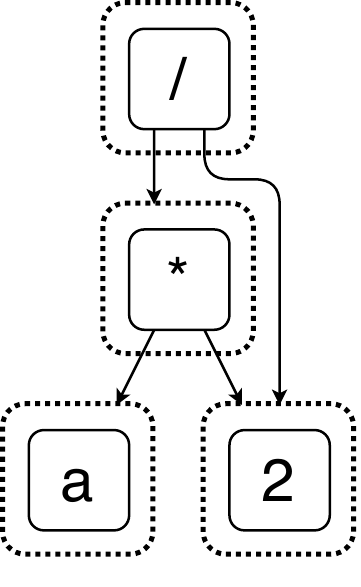}
        \caption{\Egraph represents $(a\times 2)/2$.}
    \end{subfigure}
    \hfill
    \begin{subfigure}[t]{0.30\linewidth}
        \centering
        \includegraphics[scale=0.5]{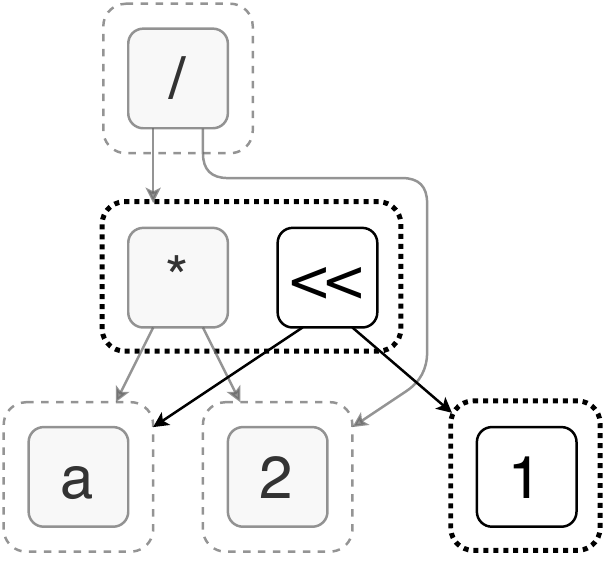}
        \caption{Rewrite \mbox{$x\times 2\rightarrow x\ll 1$}.}
    \end{subfigure}
    \hfill
    \begin{subfigure}[t]{0.30\linewidth}
        \centering
        \includegraphics[scale=0.5]{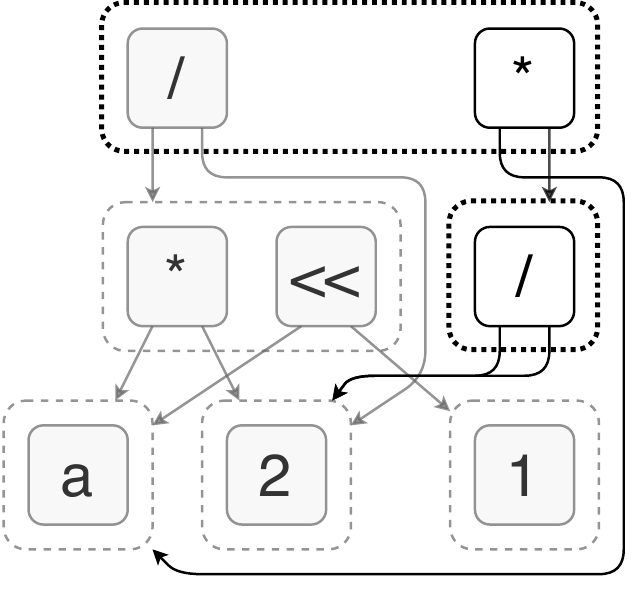}
        \caption{Rewrite $(x\times y)/z\rightarrow x\times(y/z)$.}
    \end{subfigure}
    \caption{Applying rewrites over an example \egraph (figures from \citet{egg}). 
    A solid box denotes an \enode, and a dotted box denotes an \eclass.
    \Enodes consist of a function symbol and children \eclasses, 
    and \eclasses contain a set of \enodes.}
    \label{fig:egraphs}
\end{figure}


Traditional term rewriting applies one rule at a time 
 and forgets the original term after each step, 
 so it is sensitive to the ordering of the rewrites.
For example, rewriting $(a\times 2)/ 2$ to $(a\ll 1)/2$ 
 is locally good, but it prevents future opportunities to cancel out $2/2$.
Equality saturation (\eqsat)~\cite{eqsat}
 is a technique to mitigate this
 phase-ordering problem.
\eqsat fires all the rules in each iteration 
 and keeps both original and rewritten terms in a special data structure called the \egraph.
An \egraph~\cite{nelson} is a compact data structure 
 that represents large sets of terms efficiently.
An \egraph is a set of \emph{\eclasses}, 
 and each \eclass is a set of equivalent \emph{\enodes}.
An \enode is function symbol with children \eclasses (not \enodes).

An \egraph can compactly represent 
 an exponential number of terms compared to the size of the \egraph.
We say an \egraph \emph{represents} a term $t$
 if any of its \eclasses represents $t$,
 and an \eclass represents $t$ 
 if any \enode in the \eclass represents it.
An \enode $t = f(c_1,\ldots, c_n)$
 represents a term $f(t_1,\ldots, t_n)$ 
 if each $c_i$ represents $t_i$.
An \egraph induces an equivalence relation 
 over terms: two terms are considered equivalent 
 if they are represented by the same \eclass.
This equivalence relation is also congruent: 
 if an \egraph represents two terms
 $a = f(a_1,\ldots, a_n)$ and $b = f(b_1,\ldots, b_n)$
 such that \egraph shows $a_i \equiv b_i$,
 then the \egraph can also show $a \equiv b$.%
\footnote{
  In \egraph implementations that canonicalize \enodes,
  congruence amounts to deduplication of \enodes
  since nodes $a$ and $b$ would canonicalize to identical \enodes.
}



\autoref{fig:egraphs} shows an example \egraph 
 and two rule applications.
We start with the initial \egraph representing only term $(a\times 2)/2$.
To apply a rewrite rule $x\times 2\rightarrow x\ll 1$,
 we first search for matches of left-hand patterns 
 using a procedure called \emph{\ematching} (pattern matching modulo equality).
This produces substitutions (in this case, only one: $\set{x \mapsto a}$)
 that we then we apply to right-hand side pattern.
Each resulting term (e.g., $a \times 2$) is finally merged 
 into the \eclass that the left-hand side pattern matched.


\paragraph{Extensions}
Standard \eqsat is purely syntactic.
In some cases, 
 this prevents users from writing sound rewrites.
For example,
 the rewrite $\sqrt{x^2}\rightarrow x$ 
 is sound iff $x$ is non-negative,
 but proving this requires semantic analyses.
A recent technique called \eclass analyses~\cite{egg}
 allows for semantic analyses in \eqsat.
An \eclass analysis associates every \eclass in an \egraph 
 with a semi-lattice value that is a semantic abstraction of the term.
During the \eqsat algorithm,
 the lattice data are propagated
 from children to parents \eclasses, and merged via lattice joins.
For example, 
 an analysis could track the lower bounds of \eclasses,
 which are initially $-\infty$ and increase over time
 as new terms are represented in these \eclasses via rewrites.
In \egg,
 the most popular \eqsat toolchain,
 \eclass analyses are currently limited.
An \egraph can 
 only have a single \eclass analysis,
 it can only propagate information \emph{upwards} from children to parents,
 and it requires writing low-level code in the host programming language (Rust in \egg's case).

\def\matmul{\mathit{matmul}}
\def\split{\mathit{split}}
\def\concat{\mathit{concat}} 
Multi-patterns are another commonly used extension to \ematching (and thus \eqsat).
Typically, \ematching only supports patterns matching a single term each.
A multi-pattern is a set of multiple patterns to be matched simultaneously.
For example, TenSat~\citep{tensat} is 
 an equality saturation based tensor graph optimizer
 that uses rewrite rules to share matrix multiplications.
It matches patterns $e_1 = \matmul(M_1,M_2)$ and $e_2 = \matmul(M_1,M_3)$ 
 simultaneously, 
 and then creates the expression $e_3 = \matmul(M_1,\concat(M_2,M_3))$
 and merges $e_1$ with $\split_1(e_3)$ and $e_2$ with $\split_2(e_3)$.
Previous works have developed algorithms for multi-patterns~\citep{tensat, ematching},
 but they are suboptimal and complex.

Relational \ematching~\citep{relational-ematching} 
 is a recent technique to improve \ematching performance, 
 including on multi-patterns,
 by reducing it to a relational query.
However, relational \ematching suffers from the ``dual representation'' problem.
An equality saturation engine has to 
 switch back and forth between the \egraphs and 
 the relational database representations.
This can sometimes take a significant amount of the run time,
 reducing the benefits of this approach.
Relational \ematching hints 
 at the fundamental connection between \egraphs and relational databases,
 but it only applies the insights to \ematching.
We further exploit the connection in \egglog,
 building a Datalog-inspired system that
 captures the entire \eqsat algorithm and goes beyond.


%% file: overview.tex
\section{\Egglog}
\label{sec:overview}

\Egglog is a logic programming language 
 that bears many similarities to Datalog,
 and it also incorporates features 
 that allow for program optimization and verification as in equality saturation.
In this section, 
 we approach \egglog by example,
 starting from the Datalog perspective
 and adding features until 
 it subsumes equality saturation.

\subsection{Datalog in \Egglog}

\begin{figure}
    \centering
    \hfill
    \begin{subfigure}[t]{0.40\linewidth}
        \begin{lstlisting}[language=egglog, numbers=left,escapechar=!,xleftmargin=1em]
(relation edge (i64 i64)) !\label{lst:edge-decl}!
(relation path (i64 i64)) !\label{lst:path-decl}!
            
(rule ((edge x y)) !\label{lst:tc-start}!
      ((path x y)))
(rule ((path x y) (edge y z))
      ((path x z))) !\label{lst:tc-end}!

(edge 1 2) !\label{lst:edge-1}!
(edge 2 3) !\label{lst:edge-2}!
(edge 3 4) !\label{lst:edge-3}!

(run) 
(check (path 1 4)) ;; succeeds
\end{lstlisting}
        \caption{Reachability in the classic Datalog style.}
        \label{fig:egglog-is-datalog:a}
    \end{subfigure}
    \hfill
    \begin{subfigure}[t]{0.55\linewidth}
        \begin{lstlisting}[language=egglog, numbers=left]
(function edge (i64 i64) i64)
(function path (i64 i64) i64 :merge (min old new))
            
(rule ((= (edge x y) len))
      ((set (path x y) len)))
(rule ((= (path x y) xy) (= (edge y z) yz))
      ((set (path x z) (+ xy yz))))

(set (edge 1 2) 10)
(set (edge 2 3) 10)
(set (edge 1 3) 30)

(run) 
(check (path 1 3)) ;; prints "20"
\end{lstlisting}
        \caption{Reachability including shortest path length.}
        \label{fig:egglog-is-datalog:b}
    \end{subfigure}
    \caption{
        \Egglog supports classic Datalog programs like reachability
         written in the natural way.
        Functions and \lstinline{:merge}
         allow \egglog to support Datalog with lattices similar to tools
         like Flix~\cite{flix} or Ascent~\cite{ascent}.
    }
    \label{fig:egglog-is-datalog}
\end{figure}

\Egglog uses a concrete syntax based on s-expressions,
 but despite this surface-level difference,
 readers familiar with Datalog should find many 
 \egglog programs familiar.
The program in \autoref{fig:egglog-is-datalog:a}
 computes the transitive closure of a graph,
 just like the Datalog program in \autoref{fig:datalog-tc-code}.
It first declares two relations of pairs of 64-bit integers 
 (\lstinline{i64} is one of \egglog's base types).
The \lstinline{edge} relation stores the edges of a graph 
 and is initially populated manually on lines~\ref{lst:edge-1}-\ref{lst:edge-3}.
The \lstinline{path} relation is populated by the rules on lines~\ref{lst:tc-start}-\ref{lst:tc-end}.
These rules compute the transitive closure of the \lstinline{edge} relation.
Finally, the last two lines execute the program and check that 
 there is a path from 1 to 4.

Let us take a closer look at the second rule in \autoref{fig:egglog-is-datalog:a}:
 it states that if there is a path from $x$ to $y$ 
 and an edge from $y$ to $z$, 
 then there is a path from $x$ to $z$.
In \egglog, a rule has two parts: 
 a \emph{query} and a list of \emph{actions}.%
\footnote{
Note that this is backwards from the
 more traditional Datalog syntax:
 \lstinline{path(X, Z) :- path(X, Y), edge(Y, Z).}
An \egglog rule's query and actions are analogous 
 to the body and head of a Datalog rule, respectively.
}
The query is a set of \emph{patterns},
 all of which must match for the rule to fire.
If all patterns do match, the query binds each variable to a value.
The actions dictate what happens when the rule fires,
 and they can use the variables that are bound by the query.
Typically, as in this example, the actions assert
 new facts to be added to the database.


\subsection{Functions and \lstinline{:merge}}
\label{sec:merge}

Unlike traditional Datalog, 
 \egglog stores data as partial functions rather than relations.
A relation in \egglog actually 
 desugars to a function whose return type is the built-in unit type.
To model a unary relation $R$,
 we can use a function $f_R$ to unit such that
 $f_R(x) = () \text{ if } x \in R \text{ else undefined}$.
While a Datalog program's rules add tuples to relations,
 \egglog's functions become defined for more and more tuples
 over the course of a program's execution,
 a concept that we will explore in more detail in \autoref{sec:semantics}.
Every user-defined function in \egglog is
 backed by a map (as opposed to a set in in Datalog).
Crucially, the map enforces the functional dependency from inputs to outputs.
In other words, a function maps each input to a unique output.
Throughout this paper, 
 we will use the term ``table'' to refer to either
 the backing map of an \egglog function
 or the backing set of a Datalog relation.

Consider the program in \autoref{fig:egglog-is-datalog:b}
 that computes the length of shortest path between all nodes.
Ignoring the \lstinline{:merge} declaration for now,
 the program is substantially similar 
 to the reachability program in \autoref{fig:egglog-is-datalog:a},
 but it uses functions instead of relations.
The program defines 
 \lstinline{edge} and \lstinline{path} functions to \lstinline{i64}
 rather than functions to unit (i.e., relations).
The first rule (the base case) in \autoref{fig:egglog-is-datalog:b}
 is similar to before: 
 it says that an edge of length $\mathsf{len}$ from $x$ to $y$
 implies there is a path of (at most) length $\mathsf{len}$ from $x$ to $y$.
The query uses
 \lstinline{=} to bind the output
 of the \lstinline{edge} function to the variable \lstinline{len}.
In the action, 
 we see the \lstinline{set} construct,
 which asserts
 that a function maps some arguments to a given value.
The action in the 
 analogous rule in \autoref{fig:egglog-is-datalog:a}
 desugars to \lstinline{(set (path x y) ())}.
Note that if the arguments are already mapped to a value, 
 we need to reconcile the old value with the new one
 to preserve the functional dependency.
This is resolved by the \lstinline{:merge} declaration which 
 we will describe next.

The second rule in \autoref{fig:egglog-is-datalog:b}
 is the transitive case,
 and here we see the purpose of the \lstinline{:merge} declaration.
This rule says that if there is a path from $x$ to $y$ of length $xy$,
 and an edge from $y$ to $z$ of length $yz$,
 then there is a path from $x$ to $z$ of length $xy + yz$.
But what if the function \lstinline{path} is already defined
 on the arguments $x$ and $z$?
Functions must map equivalent arguments to unique output,
 so the \lstinline{:merge} declaration tells \egglog 
 how to resolve this conflict.
Given the facts later in \autoref{fig:egglog-is-datalog:b},
 the program will discover two paths from $1$ to $3$:
 the single edge with length $30$ 
 will be discovered first,
 and then two-edge path with length $20$.
When \lstinline{(set (path 1 3) (+ 10 10))} 
 is executed,
 \egglog must come up with a single value to map \lstinline{(path 1 3)} to.
To do this,
 it evaluates the expression given after \lstinline{:merge} 
 in the function's declaration
 with \lstinline{old} and \lstinline{new} bound to the old and new values,
 respectively.
In this case,
 \lstinline{path}'s \lstinline{:merge} expression
 simply takes the minimum of the two path lengths.
It can be viewed as the join operator,
 which takes the supremum of a set of values, 
 of the min lattice over \lstinline{i64} 
 where the partial order is $x\sqsubseteq y\iff x\geq y$.
This is similar to the lattice semantics of Flix~\citep{flix},
 which also enforces functional dependency by taking the join
 over the old and new values in some lattice.
However, \egglog does not restrict the \lstinline{:merge} expression
 to only join operations over lattices.
In the following sections we will show how a \lstinline{:merge} expression
 that unifies values naturally gives rise to equality saturation.


\subsection{Sorts and Equality}

\Egglog gives the user the ability to declare new
 \emph{uninterpreted} sorts,
 and functions use these new sorts as inputs or outputs.
Crucially, values of user-defined sorts (as opposed to base types)
 can be \emph{unified} by the \lstinline{union} action.
\lstinline{union}-ing two values makes them point to the same element 
 in the underlying universe of uninterpreted sorts.
In other words, 
 values that have been unified are essentially indistinguishable
 to \egglog, 
 and all unified variables can be substituted for the same pattern variable.

\begin{figure}
    \centering
    \begin{subfigure}[t]{0.40\linewidth}
        \begin{lstlisting}[language=egglog]
(sort Node)
(function mk (i64) Node)
(relation edge (Node Node))
(relation path (Node Node))
            
(rule ((edge x y))
      ((path x y)))
(rule ((path x y) (edge y z))
      ((path x z)))

(edge (mk 1) (mk 2))
(edge (mk 2) (mk 3))
(edge (mk 5) (mk 6))

(union (mk 3) (mk 5))
(run)
(check (edge (mk 3) (mk 6)))
(check (path (mk 1) (mk 6)))
\end{lstlisting}
        \caption{Combining nodes with unification}
        \label{fig:egglog-unification:a}
    \end{subfigure}
    \begin{subfigure}[t]{0.55\linewidth}
        \begin{lstlisting}[language=egglog]
(datatype Math
  (Num i64)
  (Var String)
  (Add Math Math)
  (Mul Math Math))

;; expr1 = 2 * (x + 3)
(define expr1 (Mul (Num 2) (Add (Var "x") (Num 3)))) 
;; expr2 = 6 + 2 * x
(define expr2 (Add (Num 6) (Mul (Num 2) (Var "x"))))

(rewrite (Add a b)         (Add b a))
(rewrite (Mul a (Add b c)) (Add (Mul a b) (Mul a c)))
(rewrite (Add (Num a) (Num b)) (Num (+ a b)))
(rewrite (Mul (Num a) (Num b)) (Num (* a b)))

(run)
(check (= expr1 expr2))
\end{lstlisting}
        \caption{Basic equality saturation}
        \label{fig:egglog-unification:b}
    \end{subfigure}
    \caption{
        Unification and EqSat in \egglog.
    }
    \label{fig:egglog-unification}
\end{figure}

Consider an enhanced version of path reachability 
 in \autoref{fig:egglog-unification:a},
 where we use unification to implement node contraction
 (sometimes called vertex contraction).
The program declares a new sort \lstinline{Node},
 which is necessary because base types (like \lstinline{i64})
 cannot be unified.
The \lstinline{mk} function is the sole constructor of \lstinline{Node}s.
After the rule declarations (same as in \autoref{fig:egglog-is-datalog:a})
 and some edge assertions,
 we see our first \lstinline{union} action,
 which takes two arguments
 of the same user-defined sort and unifies them.
Now that nodes $3$ and $5$ are unified,
 running the rules will indeed find a path from $1$ to $6$, 
 a path that did not exist before the unification.

In \egglog, users define uninterpreted sorts.
A sort is a set of opaque integer values
 called \emph{ids} and an equivalence relation over those ids.
The equivalence relation is implemented
 with a union-find data structure~\cite{unionfind}
 that can \emph{canonicalize} ids;
 two ids are equivalent iff 
 they canonicalize to the same id.
Equivalent ids are considered indistinguishable by \egglog.
In fact,
 \egglog ensures that all ids appearing in the database are canonical.
These ids corresponds to \eclass ids from the \eqsat perspective.

The second line of \autoref{fig:egglog-unification:a}
 declares \lstinline{mk}, 
 a function from \lstinline{i64} to \lstinline{Node}.
This looks like a constructor, for \lstinline{Node}s, 
 but it is just like any other function from \egglog's perspective;
 the \lstinline{mk} function is backed 
 by a map from \lstinline{i64}s to \lstinline{Node} ids.
In this program,
 we never query over the \lstinline{mk} function,
 but we do call it, treating it like a total function.
What is the value of \lstinline{(mk 1)},
 especially since we did not \lstinline{set} it to anything
 prior to calling it?
Functions in \egglog can be imbued with a 
 \lstinline{:default} expression that extends
 the partial function as defined by the underlying map to be total.
Calling a function \lstinline{(f x)} will first see if
 the map for function \lstinline{f} defines 
 an output for \lstinline{x}.
If so, it returns that output.
Otherwise, 
 \egglog evaluates the \lstinline{:default} expression,
 stores the result in the map,
 and returns it.
Unless otherwise specified,
 the \lstinline{:default} for functions that output a
 user-defined sort is to create an equivalence class in the union-find
 and return its id
 (the ``make-set'' operation);
 for base types the default \lstinline{:default} 
 is to crash the program.
In other words,
 calling a function that outputs a user-defined sort
 is essentially a ``get or make-set'' operation.

The upcoming subsection will discuss how these features enable
 equality saturation,
 but \autoref{sec:points-to} will demonstrate how
 the canonicalizing union-find is useful even in a domain
 where Datalog is traditionally strong: pointer analysis.

\subsection{Terms and Equality Saturation}\label{sec:eqsat}


In \autoref{fig:egglog-unification:a},
 the \lstinline{Node} sort only has a single constructor, 
 \lstinline{mk}, which takes an \lstinline{i64}.
\egglog also supports functions that take user-defined sorts as inputs.
In this way, terms are easily constructed in \egglog.
Combined with the built-in equivalence relation,
 this term representation directly supports equality saturation in \egglog.

Consider \autoref{fig:egglog-unification:b},
 where we define a datatype \lstinline{Math} 
 that represents a simple language of arithmetic expressions.
The \lstinline{datatype} construct is sugar for
 a \lstinline{sort} declaration
 and a \lstinline{function} declaration for each constructor.
Each constructor is a function that returns a value of type \lstinline{Math},
 and its \lstinline{:default} behavior creates a fresh id
 as described above
 (we will get to its \lstinline{:merge} behavior shortly).
Now we can create terms by just nesting function calls.
The \lstinline{define} statements do just that, 
 creating two terms that we will later prove are equivalent.
These statements actually create nullary (constant) functions;
 \lstinline{(define x e)} desugars to
 \lstinline{(function x () T) (set (x) e)} 
 where \lstinline{T} is the type of \lstinline{e}.
Evaluating these terms
 adds them to the database (if not already present)
 thanks to the \lstinline{:default} behavior of the constructors.

Term rewriting in equality saturation
 has two important qualities:
 (1) pattern matching is done \emph{modulo equality} and
 (2) rewriting is \emph{non-destructive}, i.e., 
     it only adds information to the e-graph/database.
\Egglog meets both of these criteria:
 (1) all queries are performed modulo equality since
     \egglog canonicalizes the database with respect 
     to its built-in equivalence relation,
     and
 (2) \egglog rules (like standard Datalog rules) only add information
     to the database.
\Egglog provides the 
 \lstinline{rewrite} statement
 to simplify creating
 equality saturation rewrite rules.
A \lstinline{(rewrite p1 p2)} statement
 desugars to a \lstinline{rule} that queries
 for \lstinline{p1}, binds it to some variable,
 and \lstinline{union}s the variable with \lstinline{p2}:
 \lstinline{(rule ((= __var p1)) ((union __var p2)))}.

The program in \autoref{fig:egglog-unification:b}
 proves \lstinline{expr1} equivalent to \lstinline{expr2}
 using two uninterpreted rewrites
 and two that interpret the \lstinline{Add} and \lstinline{Mul} functions
 using the built-in \lstinline{+} and \lstinline{*} functions over \lstinline{i64}.
\eqsat frameworks like \egg 
 require the user to separate the uninterpreted rules
 from the interpreted part from the computed part using
 an \eclass analysis~\cite{egg}.
\egglog uses rules for both.

Like other functions that output user-defined sorts,
 the \lstinline{Math} constructors' \lstinline{:merge} behavior
 is to union the two ids.
Combined with \egglog's canonicalization,
 this means that the built-in equivalence relation is also
 a congruence relation with respect to these functions.
Consider the following map for the
 \lstinline{Add} function:
 $\{ (a, b) \mapsto c, (a, d) \mapsto e \}$.
If we \lstinline{union} $b$ and $d$ 
 such that $b$ is now canonical,
 canonicalizing the database reveals a violation 
 of the functional dependency 
 from \lstinline{Add}'s inputs to its output:
 $(a, b) \mapsto c$ but also $(a, b) \mapsto e$.
To resolve the conflict,
 \egglog invokes the \lstinline{:merge} expression
 of the \lstinline{Add} function,
 which in this case \lstinline{union}s $c$ and $e$.

After running the rules,
 the final line \lstinline{check}s 
 that \lstinline{expr1} and \lstinline{expr2} are now equivalent.
The type of both \lstinline{expr1} and \lstinline{expr2} 
 is \lstinline{Math}---a user-defined sort---%
 so the underlying value of the expressions are both ids.
Since \egglog canonicalizes the database,
 the \lstinline{check} is implemented 
 with simple equality on the ids.
\egglog supports optimization as well as verification;
 the \lstinline{extract} command prints the 
 smallest term equivalent to its given input. 


\subsection{Beyond \eqsat}

\Egglog is not limited to just Datalog or \eqsat;
 the combination allows for possibilities outside the reach of either tool.
The combining nodes example 
 from \autoref{fig:egglog-is-datalog:a}
 hints at the power of unification in Datalog,
 and \autoref{sec:points-to} takes this 
 further by implementing a unification-based pointer analysis in \egglog.
In \autoref{sec:herbie},
 we go the other way, 
 implementing several Datalog-like anaylses
 to assist an \eqsat-powered term rewriting system.

But \egglog goes beyond these applications;
 we
 describes more \egglog pearls in the full version \citep{egglog-preprint},
 including functional programming,
 type analyses for the simply typed lambda calculus in equality saturation,
 type inference for Hindley-Milner type systems,
 multivariable equational solving,
 and matrix algebra optimization with Kronecker products.
These pearls hint at the potential novel applications in program optimizations and analyses using \egglog in the future.



%% file: semantics.tex
\section{Semantics of \egglog}

In this section we describe the semantics of core \egglog.
Core \egglog differs from the full \egglog language in several aspects.
For example,
 \egglog allows multiple actions in a rule 
 while core \egglog allows only one atom in the head,
 and core \egglog does not have the \lstinline{union} operation.
These \egglog features can be desugared into the core language.
However, there are also some assumptions we made about the core \egglog.
For example, we assume the \lstinline{:merge} expression over ids
 are \lstinline{union} and the \lstinline{:merge} expressions 
 over interpreted constants is the join operator of a given lattice,
 while \egglog allows \lstinline{:merge} to be any valid \egglog expression.
In other words,
 the core \egglog captures a well-behaving subset of the full \egglog language.

\subsection{Syntax}

Given the set of (interpreted) constants $C$,
 the syntax of the core \egglog language is shown in \autoref{fig:syntax}.
An \egglog program is defined as a list of rules, 
 and each rule consists of an atom in the head
 and a list of atoms in the body.
An atom has the form $f(p_1,\ldots,p_k)\mapsto o$ 
 and intuitively means function $f$ has value $o$ on $p_1,\ldots, p_k$.
A pattern $p$ is a nested expression 
 constructed using function symbols, variables, and constants.
We additionally define a ground term (or term) $t$ to be a pattern with no variables,
 and a ground atom to be an atom where all the patterns are ground terms.

A valid \egglog program should not explicitly refer to 
 a specific uninterpreted constant $n$.
We include uninterpreted constants in the syntax nonetheless
 since they are useful when describing the semantics of \egglog programs.


\begin{figure}
  \centering
  \begin{tabular}{llcl}
     Program        & $P$ & ::= & $R_1,\ldots, R_n$                                                                 \\
     Rule           & $R$ & ::= & $A\ngets A_1,\ldots A_m$.                 \\
     Atom           & $A$ & ::= & $f(p_1,\ldots, p_k)\mapsto o\mid f(p_1,\ldots, p_k)$                 \\
     Pattern        & $p$ & ::= & $f(p_1, \ldots, p_k)\mid o$                 \\
     Term           & $t$ & ::= & $f(t_1, \ldots, t_k)\mid v$                 \\
     Base pattern          & $o$ & ::= & $v\mid x$                                               \\
     Constant       & $v$ & ::= & $c\mid n$ \\
     Interpreted Constant       & $c\in C$                                           \\
     Uninterpreted Constant       & $n\in N$                                           \\
     Variable       & $x, y, \ldots$  \\
  \end{tabular}
  \caption{
    Syntax of core \egglog. 
    }
  \label{fig:syntax}
\end{figure}

\subsection{Semantics}\label{sec:semantics}

\def\DB{\mathit{DB}}
\def\bmt{\textbf{t}}
\def\bmv{\textbf{v}}
\def\mergef{\mathit{merge}}
Given an infinite set of uninterpreted constants\footnote{
    These uninterpreted constants 
    play a similar role as \eclass ids in \eqsat
    or labelled nulls in the chase from the database literature.
}
 $N = \{n_1,n_2, \ldots\}$ 
 and a complete lattice $L = (C, \sqsubseteq, \sqcup)$ 
 over domain of interpreted constants $C$,
 We define $\bot$ to be the least element of $L$.
A schema $S$ is a collection of function symbols and their function signatures, 
 where the types range over $\{N, C\}$.
Given a schema $S$, an instance of $S$ is defined $I=(\DB, \equiv)$,
 where $\DB$ is a set of function entries $f(v_1,\ldots, v_k)\mapsto v$
 that is consistent with the schema
 and $\equiv$ is an equivalence relation over $N\cup C$ 
 satisfying $\forall c_1, c_2\in C. c_1\equiv c_2\rightarrow c_1=c_2$ 
 (i.e., interpreted constants are only equivalent to themselves).
\revision{For convenience, we also lift set operator (e.g., union, difference)
 to be between an instance and a database, 
 which applies the operator to instance's database.}

Given an arbitrary total order $<$ over $N\cup C$,
we define the canonicalization function $\lambda_{\equiv}(t)=\min t':t'\equiv t$.
For convenience, we lift $\lambda_{\equiv}$ to 
 also work on sets 
 and the whole database instances by pointwise canonicalization.

Before proceeding to define the semantics of an \egglog program,
 we need to first define what it means for a ground atom (an atom without variables)
 to be in the database and what it means to add one to the database.
First, we use the judgement $I\vdash A$ 
 to denote a ground atom is contained in the database $I$.
\begin{mathpar}
    \inferrule{
        I\vdash t_i\mapsto v_i \text{ for $i=1\ldots k$}\\\\\and
        I=(\DB{}, \equiv)\and
        f(v_1, \ldots, v_k)\mapsto v \in \DB{}
    }{
        I\vdash f(t_1, \ldots, t_k)\mapsto v
    }
    \and
    \inferrule{
    }{
        I\vdash v\mapsto v
    }
    \and
    \inferrule{
        I\vdash f(t_1, \ldots, t_k)\mapsto v \text{ for some $v$}
    }{
        I\vdash f(t_1, \ldots, t_k)
    }
\end{mathpar}

\def\flatten{\textit{flatten}}
\def\aux{\textit{aux}}

We also define $\flatten{}_I$ in \autoref{fig:flatten}
 to flatten function entries to be inserted 
 into $I$ given a nested ground atom $A$.

\begin{figure}
\begin{align*}
    \flatten{}_I(A) &= s\quad \text{where $(v, s) = \aux{}(A)$}\\
    \aux{}(f(t_1,\ldots, t_k)\mapsto v)
      & = \left(
        v,
        \{f(v_1,\ldots, v_k)\mapsto v\}\cup \bigcup_{i=1,\ldots,k}s_i
      \right)\\
      & \text{where $(v_i, s_i)=\aux{}(t_i)$ for $i=1,\ldots,k$.}\\
    \aux{}(f(t_1,\ldots, t_k))
      & = \left(
        v,
        \{f(v_1,\ldots, v_k)\mapsto v\}\cup \bigcup_{i=1,\ldots,k}s_i
      \right)\\
      & \text{where $(v_i, s_i)=\aux{}(t_i)$ for $i=1,\ldots,k$}\\
      & \text{and $I\vdash f(v_1,\ldots, v_k)\mapsto v$ 
        if such $v$ exists and $v=\textit{default}_f$ otherwise.}\\
    \aux{}(v) &= \left(v, \emptyset\right)
\end{align*}
\caption{
$\flatten{}_I(A)$ 
 flattens function entries to be inserted 
 into $I$ given a nested ground atom $A$.
If the output type of $f$ is $N$,
 then $\textit{default}_f$ is a fresh constant from $N$; 
 otherwise it is $\bot$.
The auxillary function $\aux$ 
 takes a ground atom and
 returns the ``output value'' of the ground atom
 and the set of flattened facts
 it will populate.
}\label{fig:flatten}
\end{figure}

Now we can define the semantics of an \egglog program.
It consists of two parts: the immediate consequence operator and the rebuilding operator.
We can define the (inflationary) immediate consequence
 operator $T^\uparrow_P$ as follows.%
 \footnote{
     The definition of immediate consequence operator
      in standard Datalog does not union with $\DB{}$,
     because rule applications in standard Datalog are monotone.
     This is not the case in \egglog in general.
     For example, rule $Q(e)\ngets \textit{lo}(e)\mapsto 5$, where 
      $\textit{lo}$ tracks the lower bound of an expression,
     is not monotone because the value of $\textit{lo}(e)$ can increase over time.
     Although one can adapt the meet semantics of Flix~\citep{flix} for relational joins to enforce monotonicity, 
     we do not do this in \egglog to be compatible with existing \egg applications, 
     which can be non-monotonic.
     Instead, we define \egglog semantics using the inflationary immediate consequence operator, 
     which is used to describe semantics for non-monotonic extension of Datalog such as $\text{Datalog}^\neg$~\citep{neg-by-fixpoint}.
 }
Let $\sigma$ denote a substitution that maps variables to constants,
 and let $A[\sigma]$ denote the ground atom obtained by applying $\sigma$ to atom $A$
 in the standard way.
Given an \egglog program $P$ 
 consisting of a set of rules and $I=(\DB{}, \equiv)$,
 then $T^{\uparrow}_P(I)=\DB\cup T_P(I)$ and $T_P(I)=(\DB{}', \equiv)$, where
\[
    \DB{}'=
    \bigcup_{
        (A\ngets A_1,\ldots, A_m) \in P
    }
    \left\{ 
        \displaystyle
        \textit{flatten}_I(A[\sigma])\mid
        \forall_{i=1,\ldots,m}\ 
        I\vdash A_i[\sigma]
    \right\}
\]

However, functions in $T^\uparrow_P(I)$ may no longer preserve the functional dependencies, 
 as it is possible that the same key $(v_1,\ldots, v_k)$ are mapped to more than one $v$ in some $f$.
We call $T^\uparrow_P(I)$ a \emph{pre-instance}, since it is not a valid instance yet.
To transform a pre-instance into a valid instance, 
 we further define the rebuilding operator \(
    R((\DB{}, \equiv)) = (\DB{}_R, \equiv_R)
\), where:
\begin{align*}
    (\equiv_R) &=
    \text{equivalence closure of }
    \left(
    (\equiv) \cup
    \left\{
        (n_1,n_2)\;
        \bigg|
        \begin{array}{l}
          f(v_1,\ldots, v_k)\mapsto n_1\in \DB{}, \\
          f(v_1,\ldots, v_k)\mapsto n_2\in \DB{}, \\
          n_1, n_2\in N
        \end{array}
    \right\}
    \right)
    \\
    \DB{}_R &=
    \lambda_{\equiv_R}
    \left(\left\{
        f(v_1,\ldots, v_k)
        \mapsto 
        \mergef_{f,\equiv}\left(
            K
        \right)
        \bigg|
        \begin{array}{r}
          K=\left\{v: f(v_1,\ldots, v_k)\mapsto v\in \DB{}\right\} \\
          \text{and $K$ is not empty}
        \end{array}
    \right\}\right)
    \\
    \mergef_{f,\equiv}(K) &=
    \begin{cases}
        \min\left(\lambda_{\equiv}(K)\right) &\text{if the output type of $f$ is $N$;}\\
        \bigsqcup K &\text{if the output type of $f$ is $C$.}
    \end{cases}
\end{align*}

\def\lfp{\mathit{lfp}}

Note that the canonicalization in computing $\DB{}_R$ 
 (i.e., $\lambda_{\equiv_R}$) may cause $I$ to be invalid again,
 so successive rounds of rebuilding may be needed. 
Therefore, the complete rebuilding function $R^{\infty}$ 
 is defined as iterative applications of $R$ until fixpoint.
\revision{The rebuilding process always terminates, 
 as it shrinks the size of the database in each iteration.}

We define one iteration of evaluating 
 an \egglog program $F_P$ as $R^{\infty}\circ T^\uparrow_P$, i.e.,
 do rule application once, and apply rebuilding until fixpoint. 
Intuitively, applying $F_P$ to a database makes it ``larger'', 
 in the sense that more facts may be represented.
To capture this monotonicity, we define the expanded database 
 $E_{\equiv}(\DB{})$
 such that 
 $f(v_1, \ldots, v_k)\mapsto n\in E_{\equiv}(\DB{})$ iff
  $f(\lambda_{\equiv}(v_1),\ldots, \lambda_{\equiv}(v_k))\mapsto \lambda_{\equiv}(n)\in \DB{}$
 and
 $f(v_1, \ldots, v_k)\mapsto c\in E_{\equiv}(\DB{})$ iff
  $\exists c'. f(\lambda_{\equiv}(v_1),\ldots, \lambda_{\equiv}(v_k))\mapsto c'\in \DB{}\land c\sqsubseteq c'$.
A database is larger than or equal to another database
 if it knows at least as many facts and equalities,
 so we define
 $(\DB_1, \equiv_1)\sqsubseteq_I (\DB_2, \equiv_2)$ iff
 $E_{\equiv_1}(\DB_1)\subseteq E_{\equiv_2}(\DB_2)$ 
 and
 $\equiv_1\subseteq\equiv_2$.

Although $F_P$ is not a monotone function in general,
the following sequence of iterative applications is always monotonically increasing:
\[
    I_{\bot}\sqsubseteq_I F_P(I_{\bot})\sqsubseteq_I F_P(F_P(I_{\bot}))\sqsubseteq_I  F_P(F_P(F_P(I_{\bot})))\sqsubseteq_I \ldots
\]
for initial database $I_{\bot}=(\emptyset, \textit{Id}_{N\cup C})$
where $\textit{Id}_{N\cup C}$ is the identity relation
over $N\cup C$. This ensures the existence of a fixpoint.

Finally, the result of evaluating an \egglog program $P$ is defined as the inductive fixpoint of $F_P$, 
 i.e., \(
    [\![P]\!]=F_P^{\infty}(I_{\bot})
\).
For many practical applications, 
 $[\![P]\!]$ often has an infinite size, 
 so we only calculate a finite under-approximation of the result, 
 i.e., $(R^{\infty}\circ T^\uparrow_P)^n(I_\bot)$ for some iteration size $n$. 
\revdel{
This gives the \naive algorithm for evaluating an \egglog program given iteration size $n$ shown in Algorithm~\ref{alg:eval}.
}

\def\UF{\textit{UF}}

\revision{
\subsection{Semi-na\"ive Evaluation}\label{sec:seminaive}
}


\def\SN{\mathit{SN}}
\def\N{\mathit{N}}

Last section gives an algorithm for evaluating \egglog programs, 
 which iteratively apply the immediate consequence operator ($T^\uparrow_P$)
 and the rebuilding operator $R^\infty$.
We call this algorithm the \naive evaluation.
Despite straightforward,
 the \naive evaluation may duplicate works by re-discovering same facts
 over and over again.
The \seminaive algorithm mitigates this problem 
 by incrementalizing the evaluation.
In \seminaive evaluation, each iteration maintains a differential database $\Delta\DB_i$,
 which will only contain tuples that are updated or new in this iteration.
The \seminaive rule application operator $T^\SN_P(I_i, \Delta\DB_i)$ 
 additionally takes a differential database.
For each rule $A\ngets A_1, \ldots, A_m$, $T^\SN_P$ will expand it
 into $m$ delta rules 
 $\{A \ngets A_1, \ldots, A_{j-1}, \Delta A_j, A_{j+1}, \ldots, A_m \mid j \in 1\ldots m\}$
 and apply these rules to the database similar to $T_P$.

\begin{algorithm}
    \caption{The \seminaive evaluation of an \egglog program.}
    \label{alg:eval}
    \begin{algorithmic}
    \Procedure{$F^\SN_P$}{$P, n$}
        \State $I_0\gets I_\bot$;\ \ $\Delta\DB{}_0\gets \emptyset$;
        \For{$i = 1\ldots n$}
            \State $\left(\DB_i{},\equiv_i\right) \gets R^\infty\left(
                I_{i-1}\cup T^\SN_P\left(
                    I_{i-1}, \Delta\DB_{i-1}
                \right)
            \right)$;
            \State $\Delta\DB{}_i \gets \DB{}_{i} - \DB{}_{i-1}$;
            \State $I_i\gets \left(\DB_i{},\equiv_i\right)$;
        \EndFor
        \State \Return $I_n$;
    \EndProcedure
    \end{algorithmic}
\end{algorithm}

We prove the following theorem in the full version \citep{egglog-preprint}.

\begin{theorem}
    The \seminaive evaluation of an \egglog program
    produces the same database as the \naive
    evaluation.
\end{theorem}

%% file: implementation.tex
\section{Implementation}
\label{sec:impl}

\Egglog is implemented in approximately 4,200 lines of Rust~\cite{rust}.
The codebase is open-source.\footnote{\url{https://github.com/mwillsey/egg-smol}.}
\Egglog provides both a library interface 
 and the text format shown in \autoref{sec:overview}.
As suggested by the previous sections,
 \egglog's design and implementation takes
 many cues from modern Datalog implementations~\cite{souffle,ascent,inca}.
Below, 
 we describe the design of \egglog's core components,
 as well as some of the benefits of this design.

\subsection{Components}\label{sec:components}

\Egglog's main components are 
 the database itself,
 rebuilding procedure,
 and the query engine.

\paragraph{Database}
Unlike most other Datalog implementations,
 \Egglog is based on a functional database instead of a relational database.
In other words, 
 each function/relation is backed by a map instead of a set.
This ensures that \Egglog can efficiently 
 perform the ``get-or-default'' operation required to implement terms.
For example,
 evaluating the term \lstinline{(g x)}
 will first lookup \lstinline{x} in the map for function \lstinline{g}.
If something is present, 
 it will be returned,
 otherwise \lstinline{g}'s 
 \lstinline{:default} expression is evaluated,
 placed in the map for \lstinline{(g x)},
 and returned.

The functional database also ensures that 
 a function's inputs map to a single output.
As discussed in \autoref{sec:merge},
 \egglog uses a function's \lstinline{:merge} expression
 to resolve conflicts in map.
Function conflicts can arise in two ways:
 (1) the user or a rule expressly calls 
 \lstinline{(set (f x) y)} where \lstinline{(f x)} is already defined,
 or 
 (2) a \lstinline{union} causes a function's inputs to become equivalent.
The functional database allows for efficient detection of the first case;
 the second essentially requires computing congruence closure,
 which is done by the rebuilding procedure.

\paragraph{Rebuilding Procedure}

\Egglog's rebuilding procedure is based on 
 the rebuilding procedure from \egg~\cite{egg},
 which is in turn based on the congruence closure algorithm from \citet{tarjan-congruence}.
The rebuilding procedure is responsible for
 canonicalizing the database,
 which resolves (and creates)
 the second form of function conflicts discussed above.
Suppose that a function 
 $f$ maps two different inputs to two different outputs:
 so $f(a) \mapsto b$ and $f(c) \mapsto d$.
Say that $a$ and $c$ have recently been \lstinline{union}ed,
 and that $a$ is canonical;
 rebuilding must update the entry $f(c) \mapsto d$,
 since it is no longer canonical.
Canonicalizing $f(c) \mapsto d$
 to $f(a) \mapsto d$
 causes a conflict with the previously existing entry $f(a) \mapsto b$.
To resolve conflicts,
 \egglog uses the \lstinline{:merge} expression
 to combine the two outputs into a single output.
The \lstinline{:merge} may end up \lstinline{union}ing more things,
 which may in turn create more conflicts.
The rebuilding procedure continues until
 no more conflicts are created.
For functions where the \lstinline{:merge} 
 behavior is to \lstinline{union} the two outputs,
 this is the same as congruence closure.
For other \lstinline{:merge} behavior,
 this is more akin to the \eclass analysis propagation algorithm
 from \citet{egg}.

\paragraph{Query Engine}

Once the database is canonicalized,
 \ematching is reduced to a query over the database.
Since \egglog is based on Datalog, 
 it can naturally use established techniques for efficient query execution.
\Egglog's query engine is based on 
 the relational \ematching technique from \citet{relational-ematching},
 which uses a worst-case optimal join algorithm called Generic Join~\cite{generic-join}.
\Egglog also features
 some important optimizations on top of \citet{relational-ematching}'s
 implementation
 that are only possible because of \egglog's database-native approach\footnote{
    \citet{relational-ematching}'s implementation still uses an \egraph
    data structure; 
    it creates a database from the \egraph whenever it needs to \ematch.
    \Egglog avoids this copying overhead since it is already a database.
 }, \revision{such as the \seminaive evaluation presented in \autoref{sec:seminaive}}

\subsection{Language-based Design}

Like most Datalog implementations (and \emph{unlike} most \eqsat implementations),
 \egglog is designed primarily as a programming language.
Users can write \egglog programs in a text format 
 (shown in \autoref{sec:overview}),
 and the tool parses, typechecks, compiles, and executes them.
The \egglog language includes 
 several base types (including 64-bit integers and strings)
 and operations over them.
Users can also define their own types and operations
 by using the Rust library interface.

Compared to tools like \egg
 that are more embedded in the host language,
 this design provides more
 of the user's program to the compiler for 
 checking and optimization.
For example,
 \egg provides conditional rewrite rules
 that gate a rewrite on some condition. 
The guards are essentially Rust code,
 so \egg cannot inspect them;
 it must just run the query and then check the guard.
In \Egglog, 
 there is no need for special conditional rewrites;
 all rules (and therefore rewrites) 
 can have as many conditions in the query as needed.
In addition,
 rewrite rules in \egg are not typechecked;
 \egglog prevents common errors by statically typechecking rules.
The language-based approach also allows
 the user to better model their problem
 with as many datatypes, functions, and analysis as needed.
\egg is artificially limited 
 to a single type and a single analysis in the \egraph
 due to its embedded nature;
 allowing for multiplicity would 
 significantly complicate
 the generic types in \egg's Rust implementation.

\subsection{Micro-benchmarks}

\begin{figure}[]
        \centering
        \includegraphics[width=0.5\textwidth]{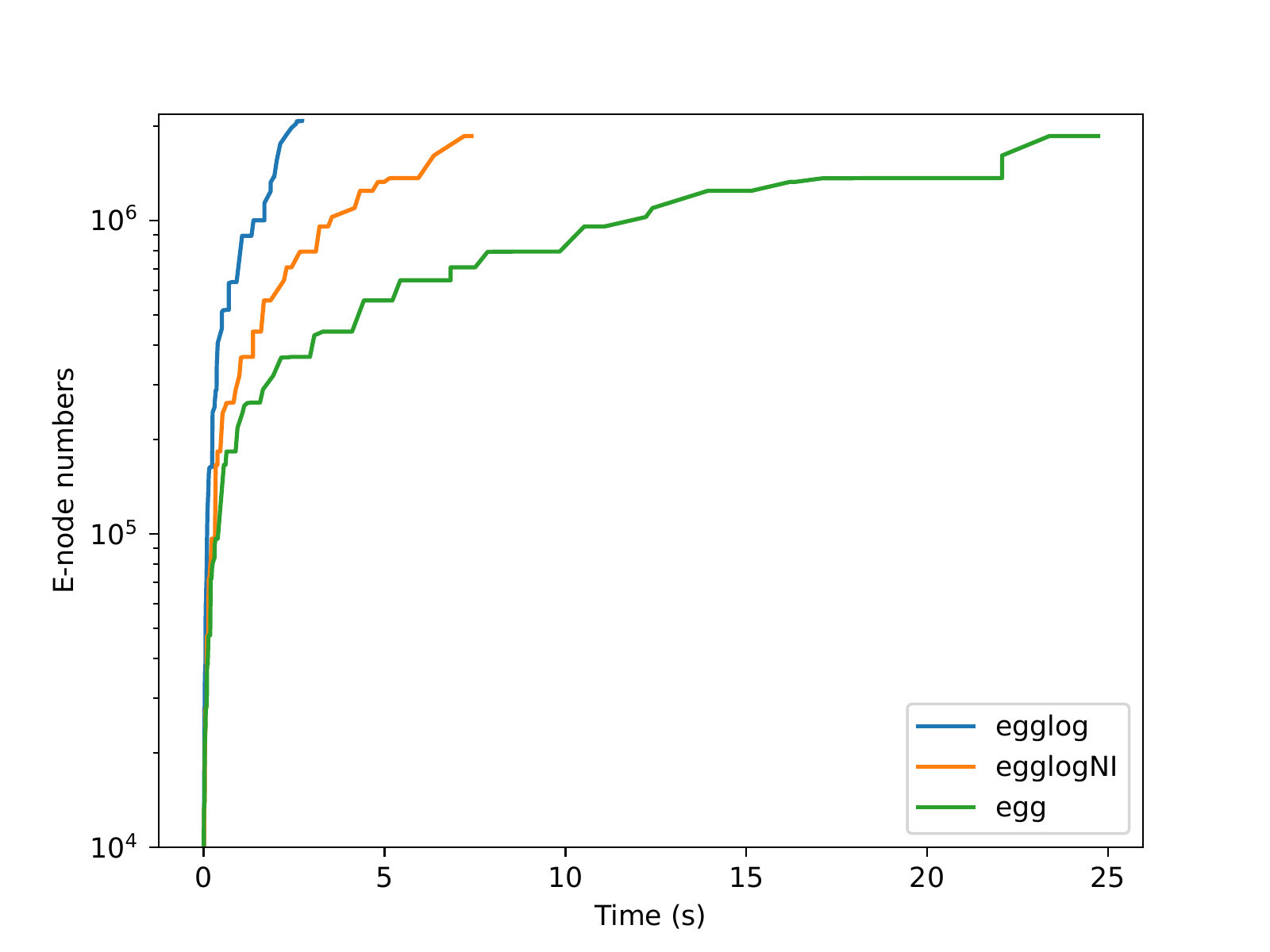}
        \caption{Performance of \egglog and \egg on \texttt{math} benchmark.
        \texttt{egglogNI} grows the same \egraph with less time than \egg.
        \egglog explores a larger program space 
        than both baselines thanks to semi-na\"ive evaluation.}
        \label{fig:microbenchmark}
\end{figure}  
In this section, we evaluate the performance of \egglog
 on a typical workload of equality saturation.
Our two baselines are \egg, 
 a state-of-the-art equality saturation framework,
 and \texttt{egglogNI}, 
 the non-incremental variant of \egglog with \seminaive evaluation disabled.
We populate all three systems with a set of initial terms 
 from \egg's \texttt{math} test suite
 and grow the \egraph with rewrite rules using \texttt{BackOff} scheduler,
 the default scheduler of \egg.
We only use a subset of rules from \texttt{math} test suite that does not involve any analysis
 (so rules that require analyses like $x/x\rightarrow 1\text{ if } x\neq 0$ are removed),
 because the scheduler behaves differently on analyses in the two systems\footnote{
    \egglog does not distinguish analyses rules from other rules,
    while \egg treats \eclass analyses specially 
    and runs them to saturation in each iteration.
 }.
As a result,
 \texttt{egglogNI} and \egg produce the same \egraph in each iteration.

We ran each system for 100 iterations.\footnote{
    All experiments in this paper are executed on 
    a MacBook Pro with Apple M2 processor and 16GB memory.
}
For each iteration, 
 we ran three systems seven times and report the median time
 versus sizes of \texttt{math} \enodes (tuples for \egglog systems).
The result is shown in \autoref{fig:microbenchmark}.
Thanks to the efficient relational matching algorithm
 and the relational query optimizer, 
 \texttt{egglogNI} grows the exact same \egraph in less time 
 than \egg, yielding a 3.34$\times$ speedup at the end of iteration 100.
Moreover,
 \egglog grows a slightly larger e-graph than \egg with a 9.27$\times$ speedup,
 because its \seminaive evaluation avoids redundant matches.

%% file: case-study.tex
\section{Case Studies}

\input{pointer-analysis.tex}

\input{herbie.tex}

%% file: pointer-analysis.tex
\subsection{Unification-Based Points-to Analysis}
\label{sec:points-to}

Many program analysis tools~\citep{bddbddb, cclyzer, doop}
 are implemented in Datalog.
The declarative nature of Datalog makes the development easier,
 and the mature relational query optimization and execution techniques
 provide competitive performance and
 sometimes lead to order-of-magnitude speedup~\citep{doop}.

Points-to analysis
 computes an over-approximation of 
 the set of possible allocations a pointer can point to at run time.
Most points-to analyses implemented in Datalog 
 are subset based (i.e., Andersen style).
These analyses are precise, 
 but they scale poorly due to its quadratic complexity.
On the other hand, 
 unification-based points-to analysis (i.e., Steensgaard style~\cite{DBLP:conf/popl/Steensgaard96})
 is nearly linear in complexity
 and therefore scales much better,
 at the cost of potential imprecision.
In a unification-based analysis,
 if it is ever learned that $p$ 
 may point to two allocations $a_1$ and $a_2$,
 the allocations are \emph{unified} and considered equivalent.
The points-to relation in a Steensgaard analysis
 is essentially a function from pointers to the equivalence class of allocations
 they point to.
This is less precise than subset-based analysis,
 but it is more scalable, 
 because it avoids tracking every individual allocation a pointer points to.

However, despite its success in hosting other program analysis algorithms, 
 classical Datalog fails to express Steensgaard analysis efficiently
 due to the lack of support for fast equivalence.
A plain representation of the equivalence relation in Datalog 
 is quadratic in space, 
 which defeats the purpose of unifying points-to allocations of a pointer.
Recently, \souffle added support for union-find--backed relations
 to benefit from the space- and time-efficient representation 
 in the union-find data structure~\citep{eqrel}.
Relations marked with the \eqrel keyword in \souffle will be stored using union-find,
 so the equivalence closure property of the relation will be automatically maintained,
 without explicit rules like transitivity, which is quadratic.
However, \eqrel relations in \souffle 
 only \emph{maintain} equivalence relations efficiently,
 but fail to interact with the rest of the rules efficiently.
Therefore,
 practical Steensgaard analyses do not use the equivalence relation directly.
Consider this rule\footnote{In contrast to our paper, 
 \citet{eqrel} views \lstinline[language=Souffle]{vpt} itself as an \eqrel relation, 
 and the body of the rule joins over only 
 \lstinline[language=Souffle]{store}, \lstinline[language=Souffle]{vpt}, \lstinline[language=Souffle]{load}.
 Our presentation here is adjusted to be consistent with \cclyzerpp.} adapted from the Steensgaard analysis benchmark 
 in the \eqrel paper~\citep{eqrel}:
\begin{lstlisting}[language=Souffle]
// *x = y; p = *q; x and q points to the same set of allocs
eql(yAlloc, pAlloc) :- store(x, y), vpt(x, xAlloc), vpt(y, yAlloc),
                       load(p, q), vpt(p, pAlloc), vpt(q, qAlloc), 
                       eql(xAlloc, qAlloc),
\end{lstlisting}
where \verb|vpt| is the points-to relation from pointers to allocations,
 and \verb|eql| is the equivalence relation declared with \eqrel.
To see the performance of this rule,
 let us consider the subquery \lstinline[language=Souffle]{vpt(x, xAlloc), vpt(q, qAlloc), eql(xAlloc, qAlloc)}.
To evaluate this subquery,
 \souffle has to join over the \lstinline[language=Souffle]{eql} relation, 
 even when it is known that \lstinline[language=Souffle]{xAlloc} and \lstinline[language=Souffle]{qAlloc} are equivalent.
We call this additional join over the equivalence relation ``join modulo equivalence''.
This occurs frequently when using equivalence relations in practice 
 and often leads to unacceptable performance.
\Egglog eliminates this join modulo equivalence 
 by actively canonicalizing each element to its canonical representation.
Because two elements are equivalent if and only if they have the same canonical representation,
 it suffices for \egglog to perform only an equality join 
 over \lstinline[language=Souffle]{vpt(x, xAlloc)} and \lstinline[language=Souffle]{vpt(q, qAlloc)} with \lstinline[language=Souffle]{qAlloc = xAlloc},
 without joining with an auxillary quadratic relation.

\cclyzerpp~\citep{cclyzerpp,cclyzer} implements Steensgaard analyses in Datalog with extensions,
 \souffle in particular.
Joins like the above are too expensive for \cclyzerpp, 
 so \cclyzerpp avoids such joins modulo equivalence as much as possible.
In fact, 
 profiling shows that 
 the only rule that involves join modulo equivalence in \cclyzerpp
 is an order of magnitude slower 
 than any other rule \cclyzerpp uses to compute the points-to analysis.

In Steensgaard analyses, 
 all allocations pointed to by the same pointer should be unified,
 so only one allocation per pointer will need to be tracked,
 which ensures an almost linear performance.
However,
 this key performance benefit 
 is lost in a direct encoding of Steensgaard analyses
 in Datalog.
For each pair of pointer \lstinline[language=Souffle]{p} 
 and the allocation it points to,
 a direct encoding will create a tuple \lstinline[language=Souffle]{vpt(p, alloc)},
 so \lstinline[language=Souffle]{vpt} may contain many allocations pointed to 
 by the same pointer, despite them all being equivalent.
The many allocations pointed to by the same pointer
 will be further propagated to other pointers,
 causing a blow up in the points-to relation.
To make sure only 
 one representative per equivalence class will be 
 propagated, \cclyzerpp uses a complex encoding
 with choice domain~\citep{choice-datalog, choice-souffle}
 and customize its own version of equivalence relation using 
 subsumptive rules~\citep{datalog-subsumption}.

Qualitatively, 
 we argue such an encoding is complex and error-prone,
 and we identify two independent bugs 
 related to the \cclyzerpp encoding.
Each bug could lead to unsound points-to analysis result.
To fix the bugs, 
 we have to bring back the \eqrel relations of \souffle.
In other words, our patched version involves 
 the interaction among choice domain, subsumptive rules, 
 and \eqrel relations, three of the newest features of \souffle.
In our experience,
 the interplay of these features can produce unexpected results
 and is extremely tricky to debug.

Compared to the sophisticated and unintuitive encoding 
 one has to develop to express efficient Steensgaard analyses
 in Datalog,
 writing Steensgaard analyses in \egglog is straightforward.
The user only needs to specify that the \lstinline[language=Souffle]{vpt} relation is 
 a function where functional dependency repair is done via 
 unifying the violating ids,
 and \egglog takes care of 
 all the unification and canonicalization.
Our insight here with \egglog is that,
 if two terms are known to be equivalent,
 they should considered indistinguishable by
 the database. 
\Egglog's canonicalization
 means we do not have to join modulo equivalence;
 a regular join suffices.

\paragraph{Benchmarking Points-To Analysis}

\begin{figure}
  \centering
  \includegraphics[width=\linewidth]{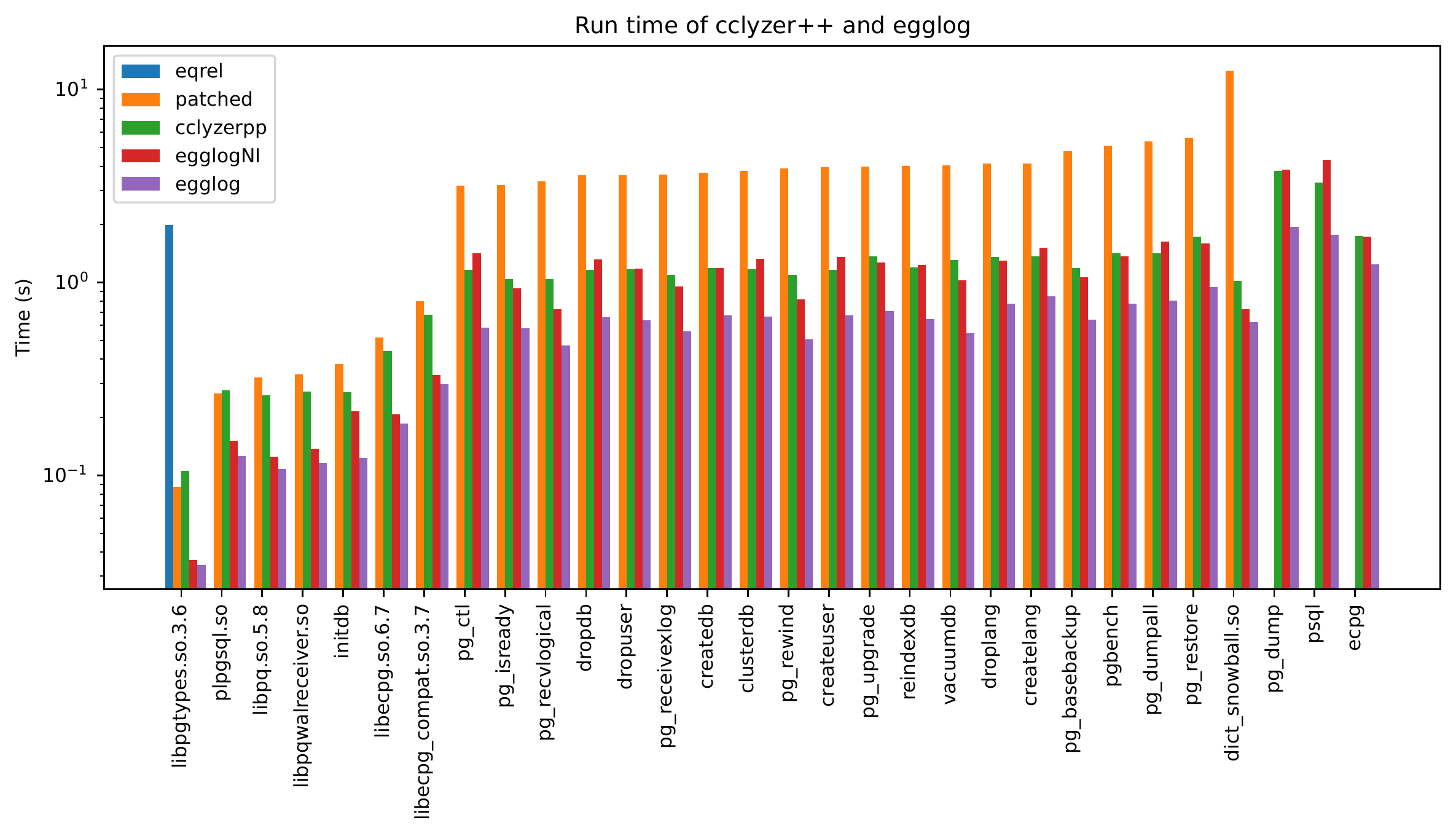}
  \caption{
    Performance comparison between \egglog and various encoding 
    of Steensgaard analyses in Souffl\'e.
    Benchmarks that time out are not shown.
    In particular, \eqrel times out on all 
    but one benchmarks and \cclyzerpp times out
    on the three benchmarks on the right.}
  \label{fig:points-to}
\end{figure}

We benchmark the performance of \egglog on points-to analyses 
 against three baselines: 
\begin{itemize}
    \item \eqrel
     uses an explicit \eqrel relation to represent
     the equivalences among allocations.
     In \eqrel, because there is no canonical representation of pointers,
      a pointer may point to multiple (equivalent) allocations in \lstinline{vpt}.
    \item \cclyzerpp uses the original encoding \cclyzerpp
     developed for Steensgaard analyses.
    It uses a custom equivalence relation 
     that keeps record of the canonical representation of an allocation
     and avoids duplicated allocations 
     with \souffle's choice domain feature.
    However, \cclyzerpp has to perform joins modulo equivalence 
     for analyzing the \lstinline{load} instruction,
     and the custom equivalence relation is semantically unsound.
   \item \texttt{patched} is a patched encoding we developed based on \cclyzerpp's encoding.
   We made the custom equivalence relation sound by bringing back the \eqrel relation in \souffle,
    while keeping the canonical representations of allocations.
   We also added an additional rule to address a congruence-related bug in \cclyzerpp.
\end{itemize}
Moreover, we compare against 
 \texttt{egglogNI}, the non-incremental variant of \egglog with semi-na\"ive disabled.

We reimplemented a subset of \cclyzerpp in \egglog and three baselines. 
The points-to analyses we implement is context-, flow-, path-insensitive and field-sensitive.
We ran points-to analyses written 
 in two variants of \egglog and the three baselines 
 on programs from postgresql-9.5.2 with a timeout of 20 seconds.
All the systems except for \cclyzerpp 
 report the same size for computed points-to relations.
\autoref{fig:points-to} shows the result.
\eqrel times out on all but one of the benchmarks, 
 and the patch to \cclyzerpp, despite making it sound,
 does make the encoding slower with the explicit equivalence relation
 and times out on three of the benchmarks.
\texttt{egglog} is faster than all Souffl\'e based baselines.
The comparison between \texttt{egglog} and \texttt{egglogNI}
 additionally shows that semi-na\"ive evaluation brings a substantial amount of 
 speedup to the computation of points-to analyses 
 by avoiding duplicated works.
Not counting the timed-out benchmarks, 
 \texttt{egglog} achieves a 4.96$\times$ speedup over \texttt{patched} on average,
 which is the fastest sound encoding in \souffle available.
Moreover, it achieves a 1.94$\times$ speedup over \cclyzerpp, 
 and 1.59$\times$ over \texttt{egglogNI}.

%% file: herbie.tex
\subsection{Herbie: Making an \eqsat Application Sound}
\label{sec:herbie}

Herbie~\cite{herbie} is a widely-used, open-source tool
 for making floating-point programs more accurate,
 with thousands of users and yearly stable releases.
Herbie takes as input a real expression,
 and returns the most accurate floating-point implementation
 it can synthesize.
Since floating-point error is a critical issue
 in scientific computing,
 Herbie is used in a variety of domains,
 including machine learning, computer graphics,
 and computational biology.

The core of Herbie's algorithm
 is to run equality saturation to explore equivalent programs.
These programs are mathematically equivalent
 over the real numbers,
 but may have different behavior over floating-point numbers.
Herbie considers candidate programs from the
 results of equality saturation,
 finding the most accurate among them.

Herbie's rewrite rules are known to be unsound,
 which has been the cause of numerous bugs in the past.
In addition, unsound rules prevent Herbie from running equality saturation
  longer once unsoundness occurs.
Unfortunately, merely removing the unsound rules makes Herbie
  useless on a large portion of its benchmark suite.
For example, \autoref{fig:herbierule} shows a rewrite rule which
  is critical to Herbie's ability to find more accurate programs.
Using these unsound rules in a sound way requires a more sophisticated
  analysis of Herbie's input programs.
This analysis was nearly impossible with Herbie's existing \egraph implementation.

\Egglog has allowed us to implement precise analyses to
 make all rewrites sound.
First, we implement an interval analysis in \egglog, allowing rules to
  utilize information about the range of terms in the program (\autoref{fig:egglogsqrt}).
This unlocks a range of crucial rules involving division, including the rule shown in \autoref{fig:herbierule2}.

\begin{figure}
  \begin{minipage}{0.40\linewidth}
  \begin{subfigure}{\linewidth}
    \[
      \frac{a*b}{c} \implies \frac{a}{\frac{c}{b}}
    \]
    \caption{
      A fraction rule which requires $b \neq 0$.
    }
    \label{fig:herbierule2}
  \end{subfigure}
  \\[1em]
  \begin{subfigure}{\linewidth}
    \[
      x - y \implies \frac{x^3-y^3}{x^2+xy+y^2}
    \]
    \caption{
      A more complex rule derived from
       the factorization of $x^3-y^3 = (x-y)(x^2+xy+y^2)$.
      This is sound if either $x \neq 0 $ or $y \neq 0$.
    }
    \label{fig:herbierule}
  \end{subfigure}
  \caption{
    Herbie~\cite{herbie} uses
    rewrite rules to create program variants
    with less floating-point error from phenomena like cancellation.
    Some rules are only sound under certain conditions.
  }
  \end{minipage}
  \hfill
  \begin{minipage}{0.55\linewidth}
  \begin{lstlisting}[language=egglog]
(function lo (Math) Rational :merge (max old new))
(function hi (Math) Rational :merge (min old new))

(rule ((= e (Sqrt a)))
      ((set (lo e) (rational 0 0))))

(rule ((= e (Sqrt a))
       (= loa (lo a)))
      ((set (lo e) (sqrt loa))))
  \end{lstlisting}
\caption{
  A few example rules for interval analysis of sqrt in \egglog.
  The $lo$ relation tracks the lower bound for each term,
   and we merge lower bounds by taking the max.
  Similarly, the $hi$ relation tracks the upper bound.
  First, we know that root of anything is non-negative.
  Next, since taking the root is monotonic,
   we can propagate the bounds from the arguments
   directly to the bounds of the result.
}
\label{fig:egglogsqrt}
  \end{minipage}
\end{figure}

While the interval analysis enables many of Herbie's rules, it is not
  sufficient for some more difficult cases.
We additionally implemented a ``not equals to'' analysis,
  which leverages both the interval analysis
  as well as facts inferred during rewriting.
\egglog's support for multiple interacting analyses
  enables cleanly separating interval and $\neq$ rules;
  in other \eqsat frameworks they would be implemented
  as a fused, monolithic analysis requiring much
  significantly more complicated rules.
The not-equals analysis allows Herbie to soundly solve one of its
  classic cancellation benchmarks: $\sqrt[3]{v+1} - \sqrt[3]{v}$.
First, the interval analysis proves that $v+1 \neq v$.
Next, the rule $a \neq b \implies \sqrt{a} \neq \sqrt{b}$
 implies $\sqrt[3]{v+1} \neq \sqrt[3]{v}$.
This allows us to finally apply
  the rewrite from \autoref{fig:herbierule},
  substituting $\sqrt[3]{v+1}$ for $x$ and $\sqrt[3]{v}$ for $y$,
  reducing the error of the expression from extremely high to near zero.

\begin{figure}
    \includegraphics[scale=0.5]{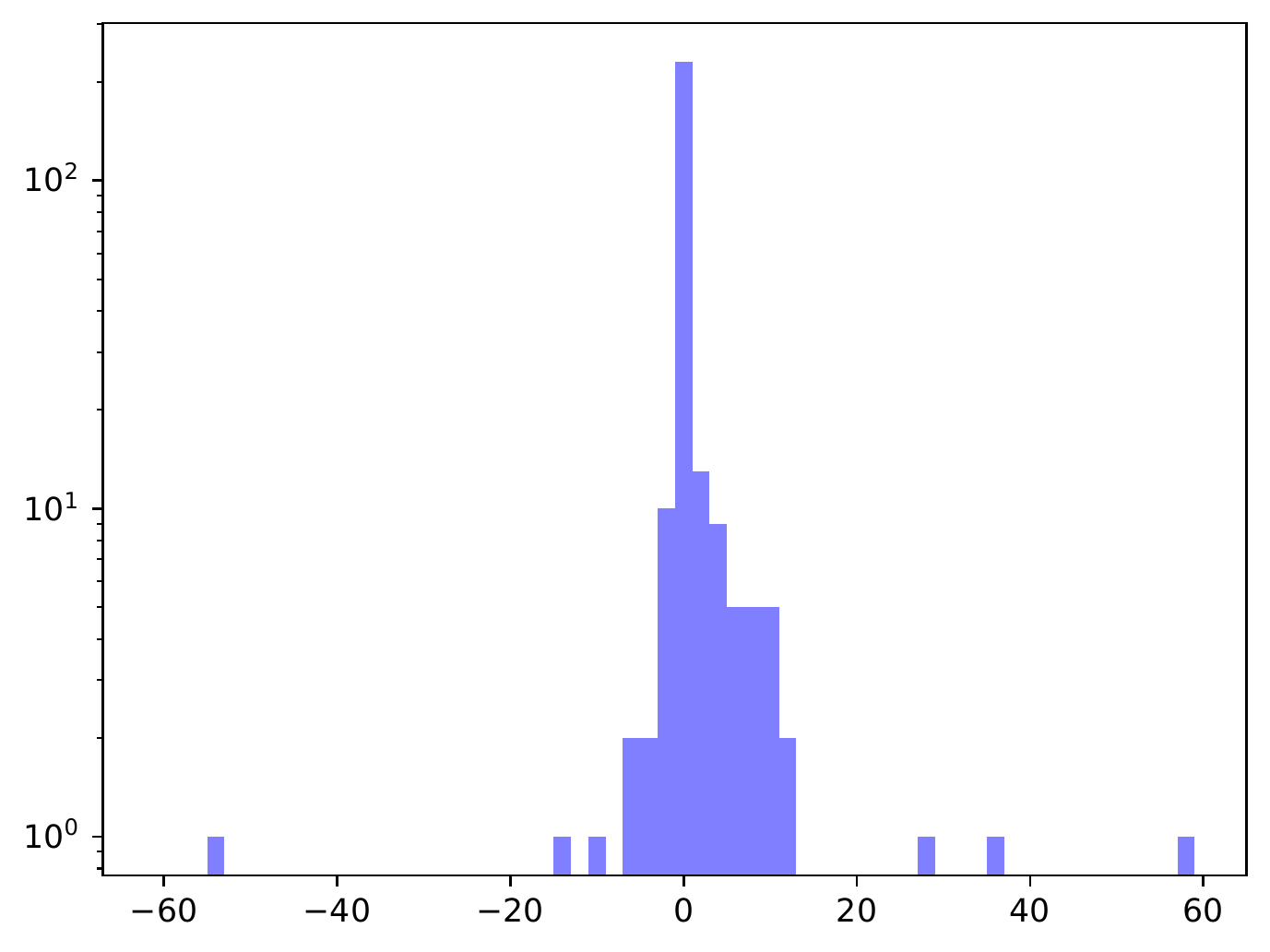}

  \caption{Graph showing the difference in error between Herbie using
    \egglog's sound analysis and Herbie using the unsound ruleset across all of Herbie's benchmark suite.
  The horizontal axis is the difference 
   in the average bits of error using 
   Herbie's unsound rules vs.\ \egglog's sound rules.
  The vertical axis is the number of benchmarks. 
  Negative values
    represent benchmarks in which Herbie found a more accurate program using \egglog's analysis.}
  \label{fig:herbieeval}
\end{figure}

\begin{figure}
  \includegraphics[scale=0.5]{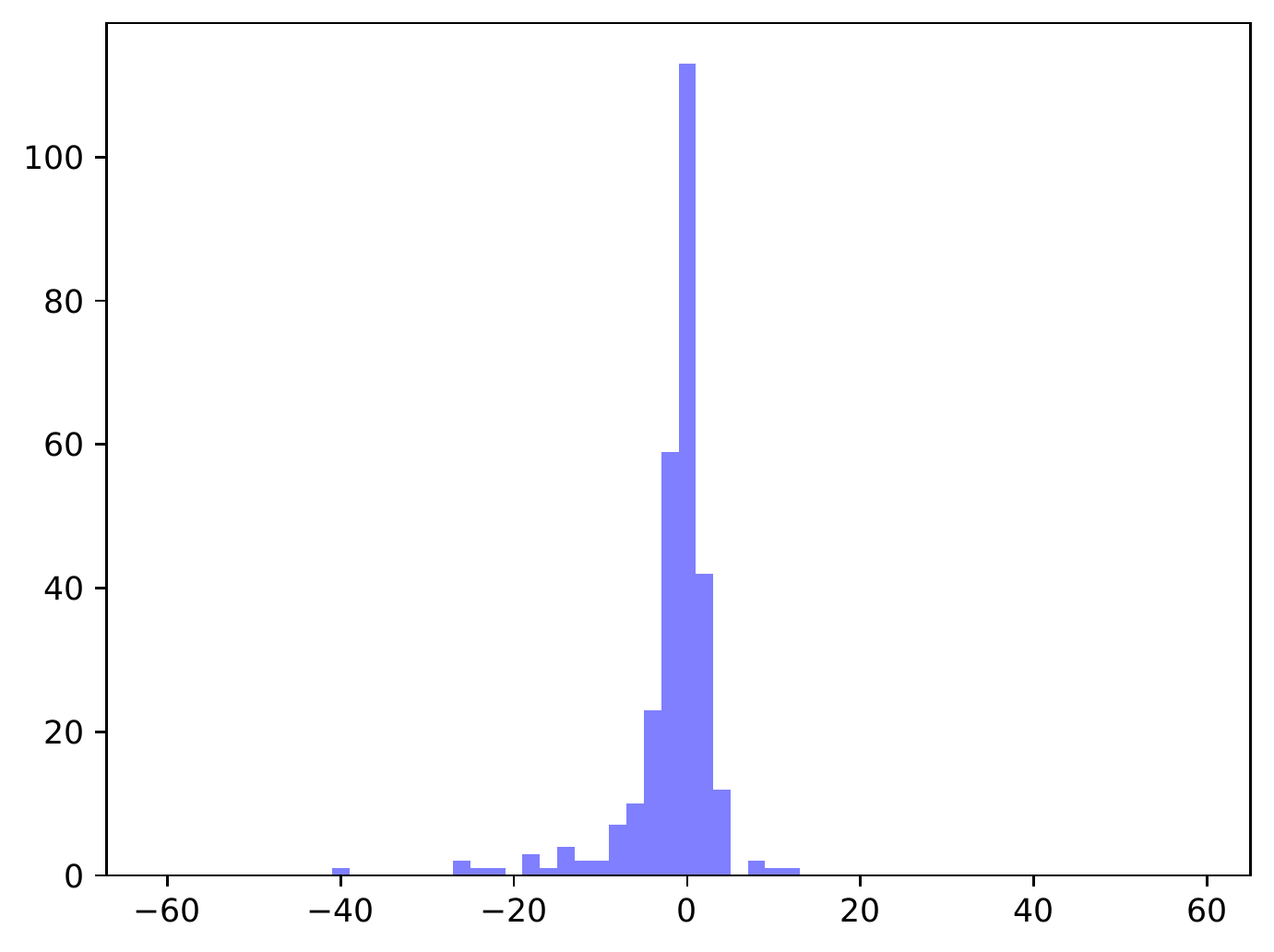}
  \caption{Graph showing the difference in runtime (in seconds)  between Herbie using \egglog's sound analysis and Herbie using the unsound ruleset across all of Herbie's benchmark suite.
  The horizontal axis is the difference in time to execute the benchmark.
  The vertical axis is the number of benchmarks.
  Negative values represent benchmarks in which Herbie was faster using \egglog's analysis.}
  \label{fig:herbietime}
\end{figure}

\autoref{fig:herbieeval} shows the results
 of our evaluation of Herbie using \egglog.
Herbie has a benchmark suite of \numtests floating-point programs,
 collected from a variety of domains.
We ran Herbie on each of these programs
 using both the unsound ruleset and \egglog's sound analysis.
Using \egglog's analysis makes Herbie faster overall
  (\timeegglogminutes minutes vs \timevanillaminutes minutes).
This is because \egglog generates no unsound programs, which slow down
  Herbie's search.

In \numbettererror cases,
 Herbie using a sound analysis is actually able
 to find a more accurate program than Herbie
 using the unsound ruleset.
In \numworseerror cases,
 Herbie's unsound ruleset finds more accurate results
 than \egglog's sound analysis.
There are several outliers in \autoref{fig:herbieeval}.
The point on the far left represents a benchmark
  which Herbie using the unsound ruleset is unable to solve.
This input program is $9x^4 - y^2(y^2-2)$,
  and the solution involves an algebraic rearrangement
  and fma (fused multiply-add) operation.
The point on the far right represents a program that
  overflows \egglog's rational type, which can be easily fixed in the future.

%% file: relatedwork.tex
\section{Related Work}

\paragraph{E-graphs and Equality Saturation}
\Egraphs were first introduced by \citet{nelson}
 in late 1970s 
 to support a decision procedure for the theory of equalities.
\citet{tarjan-congruence} later introduced
 a more efficient algorithm
 and analyzed the its time complexity.
\Egraphs are used at the core of 
 many theorem provers and solvers~\citep{simplify, z3, cvc4}.
Because \egraphs can compactly represent program spaces,
 they were repurposed for program optimization in the 2000s~\citep{eqsat,denali}.
Other data structures for compact program space representations are developed in parallel,
 including finite tree automata~\citep{dace,blaze} and version space algebras~\citep{vsa,flashmeta}.
There are two essential problems to these data structures:
 how to construct the desired program space and how to search it.
\citet{eqsat} observed that 
 the program space can be grown via equational, non-destructive rewrites, 
 which they called equality saturation.
This insight
 leads to a line of work on using equality saturation 
 for program optimization and program synthesis~\citep{herbie,ruler,tensat,spores,diospyros,szalinski}.
However, 
 a problem with this rewriting-based approach to program space construction is that,
 in many cases,
 sound rewrite rules are difficult to define in a purely syntactic way.
The \egg framework by \citet{egg} mitigates this issue 
 by introducing \eclass analyses, 
 which allow simple semantic analyses over the \egraphs.
Our work improves \eclass analyses 
 by allowing more expressive analysis rules to be defined
 compositionally.

\citet{relational-ematching} first studied 
 the connection between \egraphs and relational databases.
By reducing pattern matching over \egraphs to relational queries,
 they made the matching procedure orders of magnitude faster.
However, their technique has the dual representation problem, 
 i.e., one has to keep both the \egraph and its relational representation, 
 which limits its practical adoptions.
We build on their work
 and view the entire equality saturation algorithm from the relational perspective.
This saves us from synchronizing two representations of \egraphs
 and further exploits the performance benefits of the relational approach.

\Egglog also brings new insights on some problems in \egraphs and equality saturation.
For example, 
 the literature studied the problem of incremental pattern matching over \egraphs.
\citet{relational-ematching} conjectured that this problem can be solved by 
 classical techniques of incremental view maintenance in databases.
We complement their argument with a concrete implementation of 
 incremental matching using semi-na\"ive evaluation~\citep{seminaive}.
Moreover, 
 the literature studied the ``proof'' problem on \egraphs~\citep{pp-congr,flatt2022small}:
 many domains require not only the optimized terms that are equivalent to the original terms,
 but also a proof why they are equivalent.
\revdel{The database counterpart of this problem, called provenance, is also studied by database researchers~\citep{prov-semiring,prov-souffle}.}
A future direction is to study proof generations for \egglog programs%
\revdel{~by incorporating algorithms developed in both fields}.


\paragraph{Datalog and Relational Databases}
Functional dependency repairs via lattice joins in \egglog is directly inspired by Flix~\citep{flix}.
Flix extends Datalog by allowing relations to be optionally annotated by lattices.
With this feature, 
 Flix is able to express many advanced program analyses algorithms efficiently.
\egglog can simulate Flix programs by setting \merge to the lattice join operator.   
Flix can be regarded as among the works that 
 try to find a theoretical foundation for recursive aggregates~\citep{agg-semantics,datalog-subsumption,mono-agg,datalogo}.
A similar lattice-based approach to recursive aggregates is studied by $\text{Bloom}^L$~\citep{bloom-lattice}.
\revision{Other Datalog systems that support recursive aggregates include LogicBlox~\citep{logicblox} 
 and Rel~\citep{rel-refernce}.
}

Rewrite rules in \egglog generalize rules in Datalog 
 because the heads of rules can generate fresh ids.
This is called tuple-generating dependencies (TGDs) in the database literature.
Moreover, 
 while functional dependencies has the form $R(x_1,\ldots,x_i,\ldots,x_k),R(x_1,\ldots,x_i',\ldots x_k)\rightarrow x_i=x_i'$
 equality-generating dependencies (EGDs) generalize functional dependencies 
 by allowing equalities between different columns in different relations.
A family of algorithms called the chase can be used to reason about both TGDs and EGDs~\citep{chase-revisited,bench-chase}.
Moreover,
 the model semantics of the chase directly 
 gives a model semantics of a subset of \egglog where 
 union is the only \lstinline[language=egglog]{:merge} operation.
Compared to general TGDs and EGDs,
 \egglog imposes syntactic constraints over the programs 
 so that rules applications are deterministic and efficient.
$\text{Datalog}^{\pm}$~\citep{datalogpm} is 
 a family of extensions to Datalog based on TGDs and EGDs for ontological reasoning.

\revision{
Concurrent to our work and independent to the work on the chase, 
 \citet{bidlingmaier2023algebraic}
 formalizes Datalog with equality, which shares the same core idea to \egglog, 
 as relational Horn logic and partial Horn logic
 and studies its properties from a categorical point of view.
\citet{bidlingmaier2023evaluation} further describes an evaluation algorithm 
 for Datalog with equality similar to the chase.
Different from theirs, our work is motivated by practical applications 
 in program optimization and program analysis,
 and we focus on a simpler operational model of \egglog.
}

\revision{
While termination for Datalog with a variety of extensions is well studied, 
 the termination condition of \egglog is quite open.
In the future, 
 we hope to better understand the termination of \egglog by 
 further studying the connection between \egglog and the chase.
For example, 
 the database theory community has established many conditions 
 for chase termination (e.g., \citet{data-exchange,datalogpm,vadalog}),
 and one could potentially apply these results to \egglog
 by translating \egglog rewrite rules and functional dependencies into TGDs and EGDs.
On the other hand,
 nearly all instantiations of equality saturation in practice will diverge . 
Being a generalization of equality saturation, \egglog allows for divergence by design.}

\revision{
A key feature of \egglog is its efficient equational reasoning.
Although equational reasoning can be expressed in Datalog 
 with an explicit equivalence relation,
 doing so is very inefficient.
The \texttt{patched} baseline in \autoref{sec:points-to} 
 is a slightly more efficient encoding of equational reasoning in Datalog
 using \souffle extensions including choice domain and subsumptive rules.
We also attempted several other approaches to optimize equational reasoning 
 with existing features such as recursive aggregates and
 Constraint Handling Rule's simpagation rules \citep{chr}.
However, 
 we found none of these encodings 
 provide a natural abstraction nor competent performance.
}


\paragraph{Logic Programming and Automated Theorem Proving}
\revdel{
\egglog bears some similarity with logic programming languages like Prolog.
Prolog has unification variables similar to \egglog, except that they are backtrackable.
In general, Prolog allows disjunction and backtracking search
 at the cost of incomplete search strategy and potentially getting into infinite loops.
Moreover, Prolog programs are usually evaluated top-down,
 while Datalog and \egglog programs are evaluated bottom-up.
}

\revision{
\egglog bears some similarity with logic programming languages like Prolog.
Similar to unification variables in Prolog, 
 fresh ids in \egglog can represent unknown information 
 (see, e.g., \citet[Appendix~A.1]{egglog-preprint}), 
 and the congruence closure can be viewed 
 as a dual procedure to unification \citep{congr-duality}. 
In \citet[Appendix~A.3]{egglog-preprint}, we also show an 
 implementation of a Hindley-Milner type inference algorithm,
 of which the key construct is the unification mechanism implemented 
 as a few \egglog rules.

However, 
 several distinguishing features make \egglog highly efficient for its target domains, namely program analysis and optimization. 
For example,
 \egglog does not allow backtracking, 
 so its union-find data structure does not need to be backtrackable or persistent (unlike in Prolog or SMT solvers), 
 which makes it efficient for tasks that are monotone in nature 
 (e.g., equality saturation and pointer analysis).
Moreover, \egglog uses a bottom-up evaluation algorithm 
 more similar to Datalog than Prolog 
 (top-down backtracking search). 
One way of (partially) viewing \egglog is as a logic programming language that combines the bottom-up evaluation of Datalog
 and the unification mechanism of Prolog. The ``magic-set transformation'' is a closely related technique to simulate top-down evaluation 
 in a bottom-up language in a demand-driven fashion. 
We show in the full version \citep{egglog-preprint} several pearls that uses this idea to simulate top-down evaluations.
On the other hand,
 Prolog has imperative features such as \texttt{cut}, which removes choice points.
 \Egglog does not have a direct analog of \texttt{cut}
 (because \egglog does not backtrack), 
 although \egglog has other imperative features borrowed from EqSat techniques
 that makes fine control of the execution such as rule scheduling.
}

SMT solvers are powerful tools for deciding combinations of logic theories and
 automatically proving theorems~\citep{z3,cvc4}.
Many rewrite rules in \egglog can be expressed as SMT axioms.
In fact,
 SMT solvers support a richer language than \egglog with features like disjunction 
 and built-in theories like the theory of integer programming.
However, a key difference between \egglog and SMT solvers is that
 the output of \egglog is \emph{minimal} (or universal in database terminology).
While it is possible to ``hack into'' an SMT solver to repurpose it 
 as an EqSat engine~\cite{flatt2022small}, 
 such techniques are arcane and not officially supported.
\eqsat and \egglog's native support for extraction
 makes them better suited for program optimization.
Recently, researchers have extended Datalog to dispatch more complex constraints 
 to be solved by an SMT solver~\cite{DBLP:journals/pacmpl/Bembenek0C20}.
This greatly extends the reach of Datalog, 
 allowing the user to specify constraints in a variety of theories and logics.
In \egglog we emphasize efficiency, 
 and chose more conservative extensions that can be implemented 
 by fast data structures like union-find.

%% file: conclusion.tex
\section{Conclusion}

\egglog unifies both Datalog and \eqsat
  style fixpoint reasoning.
From the perspective of a Datalog programmer,
  \egglog adds fast and extensible equivalence relations
  that still support key database optimizations
  like query planning and semi-naive evaluation.
From the perspective of an \eqsat user,
  \egglog adds composable analyses,
  extensible uninterpreted functions, and
  incremental e-matching,
  thus significantly simplifying
  complex conditional rewrites and
  scalable program analyses.
\egglog's novel \textit{merge expressions}
  for user-specified functional dependency repair
  are the key technical mechanism enabling this
  synthesis of fixpoint reasoning capabilities.


%% file: pearls.tex
\section{\Egglog by Example}\label{sec:pearls}

In this section, we will walk through a list of examples
 showing the wide applications of \egglog.

\subsection{Functional Programming with \Egglog}\label{sec:funprog}

 \begin{figure}
     \centering
     \begin{subfigure}[t]{0.45\linewidth}
         \centering
 \begin{lstlisting}[language=Souffle]
 .type Tree = Leaf {} | Node {t1: Tree, t2: Tree}
 .decl tree_size_demand(l: Tree)
 .decl tree_size(t: Tree, res: number)
 // populate demands from roots to leaves
 tree_size_demand(t1) :-
     tree_size_demand($Node(t1, t2)).
 tree_size_demand(t2) :-
     tree_size_demand($Node(t1, t2)).
 // calculate bottom-up
 tree_size($Node(t1, t2), s1 + s2) :-
     tree_size_demand($Node(t1, t2)),
     tree_size(t1, s1),
     tree_size(t2, s2).
 tree_size($Leaf(), 1).
 // compute size for a particular tree
 tree_size_demand($Node($Leaf(), $Leaf())).
         \end{lstlisting}
         \caption{\texttt{tree\_size} in \souffle}
         \label{fig:treesize:souffle}
     \end{subfigure}
     \hfill
     \begin{subfigure}[t]{0.45\linewidth}
         \begin{lstlisting}[language=egglog]
 (datatype Tree (Leaf) (Node Tree Tree))
 (datatype Expr (Add Expr Expr) (Num i64))
 (function tree_size (Tree) Expr)
 ;; compute tree size symbolically
 (rewrite (tree_size (Node t1 t2))
     (Add (tree_size t1) (tree_size t2)))
 ;; evaluate the symbolic expression
 (rewrite (Add (Num n) (Num m))
     (Num (+ n m)))
 (union (tree_size (Leaf)) (Num 1))
 ;; compute size for a particular tree
 (define two (tree_size (Node (Leaf) (Leaf))))
         \end{lstlisting}
         \caption{\texttt{tree\_size} in \Egglog}
         \label{fig:treesize:egglog}
     \end{subfigure}
     \caption{Computing tree size with \souffle and \egglog}
 \end{figure}


 \Egglog is capable of evaluating many functional programs very naturally.
 The standard evaluation of Datalog programs is traditionally done bottom up.
 Starting from the facts known,
  it iteratively derives new facts,
  and when it terminates,
  all the derivable facts
  are contained in the database.
 In the evaluation of functional programs,
  however,
  we usually start with the goal of evaluating a big expression and
  break it down into smaller goals while
  traversing the abstract syntax tree from root to leaves,
  before collecting the results bottom up.
 In order to simulate the top-down style of evaluation
  of functional programs,
  Datalog programmers need to create manual ``demand'' relations
  that carefully tune the firing of rules
  to capture the evaluation order of functional programs.
 On the other hand,
  \egglog express many functional programs very naturally,
  thanks to the unification mechanism.

 For example, consider the task of computing the relation \verb|tree_size|,
  which maps trees to their sizes.
 A full instantiation of the \verb|tree_size| finds the size of \emph{all} trees
  and therefore is infinite,
  so bottom-up evaluations will not terminate
  in languages like \souffle.
 We need to manually \textit{demand transform} the program
  to make sure we only instantiate \verb|tree_size|
  for the trees asked for and their children.
 Demand transformation first populates a ``demand'' relation,
  and \verb|tree_size| will compute only trees that resides in the demand relation.
 The program is shown in \autoref{fig:treesize:souffle}.
 To get the size of a specific tree,
  we have to first insert the tree object into the \verb|tree_size_demand|
  relation to make a demand, before looking up \verb|tree_size| for
  the actual tree size.

 Similar to Datalog, \egglog programs are evaluated bottom up.
 However, we do not need a separate demand relation in \egglog,
  because we can use ids
  to represent unknown or symbolic information.
 To query the size of a tree \verb|t|,
  we simply put the atom \verb|(tree_size t)| in the action.
 \Egglog will create a fresh id as a placeholder for the value
  \verb|tree_size| maps to on \verb|t|,
  and the rest of the rules will
  figure out the actual size subsequently.
 The \egglog program is shown in \autoref{fig:treesize:egglog}.

 Conceptually, we create a ``hole'' for the value
  \verb|(tree_size t)| is mapped to.
 A series of rewriting will utlimately fill in this hole
  with concrete numbers.
 We use fresh ids here in a way that is similar to
  how logic programming languages use logic variables.
 In logic programming,
  logic variables represent unknown information
  that has yet to be resolved.
 We view this ability to represent the unknown
  as one of the key insights
  to both \egglog and logic programming languages
  like Prolog and miniKanren.
 However, unlike Prolog and miniKanren,
  \egglog does not allow backtracking
  in favor of monotonicity and efficient evaluations.

\subsection{Simply Typed Lambda Calculus}\label{sec:stlc}


Previous equality saturation applications use lambda calculus
 as their intermediate representation for program rewriting
~\citep{koehler2021sketch,egg,storel}.
To manipulate lambda expressions in \egraphs, a key challenge is
 performing capture-avoiding substitutions,
 which requires tracking the set of free variables.
A traditional equality saturation framework will represent
 the set of free variables
 as an \eclass analysis
 and uses a user-defined applier to perform the capture-avoiding substitution,
 both written in the host language (e.g., Rust).
As a result,
 users need to reason about both rewrite rules and custom Rust code
 to reason about their applications.

We follow the \texttt{lambda} test suite of \egg \citep{egg}
 and replicate the lambda calculus example in \egglog.
Instead of writing custom Rust code for analyses,
 we track the set of free variables using standard \egglog rules.
Figure \ref{fig:free-var} defines a function that maps terms to set of variable names.
Since the set of free variables can shrink in equality saturation
 (e.g., rewriting $x-x$ to $0$ can shrink the set of free variables from $\{x\}$ to the empty set),
 we define the merge expression as set intersections.
The first two rules say that values have no free variables and variables have themselves
 as the only free variable.
The free variables of lambda abstractions, let-bindings, and function applications are
 inductively defined by constructing the appropriate variable sets at the right hand side.
Finally,
 the last three rewrite rules perform the capture-avoiding substitution
 over the original terms depending on the set of free variables.
When the variable of lambda abstraction is contained in the set of free variables
 of the substituting term,
 a new fresh variable name is needed.
We skolemize the rewrite rule
 so that the new variable name is generated deterministically.
Note that the last two rules depend both positively and negatively
 on whether the set of free variables contains a certain variable,
 so this program is not monotonic in general.


\Egglog can not only express \eclass analyses,
 which are typically written in a host language like Rust,
 but also semantic analyses not easily expressible in \eclass analyses.
For example, consider an equality saturation application
 that optimizes matrix computation graphs and uses lambda calculus
 as the intermediate representation.
Users may want to extract terms with the least cost as the outputs of optimizations,
 but a precise cost estimator may depend on the type and shape information of an expression
 (e.g., the dimensions of matrices being multiplied).
Expressing type inference within the abstraction of \eclass analyses is difficult:
 in \eclass analyses, the analysis values are propagated bottom up,
 from children to parent \eclasses.
However, in simply typed lambda calculus,
 the same term may have different types depending on the typing context,
 so it is impossible to know the type of a term without first knowing the typing context.
Because the typing contexts need to be propagated top down first,
 \eclass analysis is not the right abstraction for type inference.
In contrast,
 we can do type inference in \egglog
 by simply encoding the typing rule for simply typed lambda calculus in a Datalog style:
 we first break down larger type inference goals into smaller ones,
  propagate demand together with the typing context top down,
  and assemble parent terms' types based on the children terms' bottom up.

\autoref{fig:stlc} shows a subset of rules that perform type inference
 over simply typed lambda calculus.
We determine the types of variables based on contexts.
For lambda expressions, we rewrite the type of \lstinline{(Lam x t1 e)}
 to be $t_1\rightarrow t_2$, where $t_2$ is the type of $e$ in the new context
 where $x$ has type $t_1$ (i.e., \lstinline{(typeof (Cons x t1 ctx) e)}).
Finally, because we cannot directly rewrite the type of function applications
 in terms of types of their subexpressions,
 we explicitly populate demands for subexpressions and
 derive the types of function applications using
 the types of subexpressions once they are computed.

 \begin{figure}
    \begin{minipage}[t]{0.48\linewidth}
\begin{lstlisting}[language=egglog, escapeinside={(*}{*)}]
(function free (Term) (Set Ident)
    :merge (set-intersect old new))

;; Computing the set of free variables
(rule ((= e (Val v)))
      ((set (free e) (empty))))
(rule ((= e (Var v)))
      ((set (free e) (set-singleton v))))
(rule ((= e (Lam var body))
       (= (free body) fv))
      ((set (free e) (set-remove fv var))))
(rule ((= e (Let var e1 e2))
       (= (free e1) fv1) (= (free e2) fv2))
      ((set (free e) (set-union fv2
          (set-remove fv1 var)))))
(rule ((= e (App e1 e2))
       (= (free e1) fv1) (= (free e2) fv2))
      ((set (free e) (set-union fv1 fv2))))

;; [e2/v1](*$\lambda$*)v1.e1 rewrites to (*$\lambda$*)v1.e1
(rewrite (subst v e2 (Lam v e1))
         (Lam v body))
;; [e2/v2](*$\lambda$*)v1.e1 rewrites to (*$\lambda$*)v1.[e/v2]e1
;; if v1 is not in free(e2)
(rewrite (subst v2 e2 (Lam v1 e1))
         (Lam v1 (subst v2 e2 e1))
    :when ((!= v1 v2)
           (set-not-contains (free e2) v1)))
;; [e2/v2](*$\lambda$*)v1.e1 rewrites to (*$\lambda$*)v3.[e/v2][v3/v1]e1
;; for fresh v3 if v1 is in free(e2)
(rule ((= expr (subst v2 e2 (Lam v1 e1)))
       (!= v1 v2)
       (set-contains (free e2) v1))
      ((define v3 (Skolemize expr))
       (union expr (Lam v3 (subst v2 e2
           (subst v1 (Var v3) e1))))))
\end{lstlisting}
\caption{Free variable analysis
 and capture avoiding substitution in \egglog.
We use skolemization function \lstinline{Skolemize}
 to deterministically generate fresh variables
 for capture-avoiding substitution.
}
\label{fig:free-var}
\end{minipage}
\hfill
\begin{minipage}[t]{0.48\linewidth}
\begin{lstlisting}[language=egglog]
(function typeof (Ctx Expr) Type)

(function lookup (Ctx Ident) Type)
(rewrite (lookup (Cons x t ctx) x) t)
(rewrite (lookup (Cons y t ctx) x)
         (lookup ctx x)
    :when ((!= x y)))

;; Type of matrix constants
(rewrite (typeof ctx (fill (Num n) (Num m) val))
         (TMat n m))

;; Type of variables
(rewrite (typeof ctx (Var x) )
         (lookup ctx x))

;; Type of lambda abstractions
(rewrite (typeof ctx (Lam x t1 e))
         (Arr t1 (typeof (Cons x t1 ctx) e)))

;; Populate type inference demand for
;; subexpressions of function application
(rule ((= (typeof ctx (App f e)) t2))
      ((typeof ctx f)
       (typeof ctx e)))

;; Type of function application
(rule ((= (typeof ctx (App f e)) t)
       (= (typeof ctx f) (Arr t1 t2))
       (= (typeof ctx e) t1))
      ((union t t2)))
\end{lstlisting}
\caption{Type inference for simply typed lambda calculus
 with matrices.
 Here we require that lambda abstractions are annotated with parameter types.
 Section~\ref{sec:hm} looses this restriction.}
\label{fig:stlc}
\end{minipage}
\end{figure}

\subsection{Type Inference beyond Simply Typed Lambda Calculus}\label{sec:hm}

\Egglog is suitable for expressing a wide range of unification-based algorithms
 including equality saturation (Section~\ref{sec:eqsat}) and Steensgard analyses (Section~\ref{sec:points-to}).
In this section, we show an additional example on the expressive power of \egglog:
 type inference for Hindley-Milner type systems.
Unlike the simple type system presented in Section~\ref{sec:stlc},
 a Hindley-Milner type system does not require type annotations for variables in lambda abstractions
 and allows let-bound terms to have a \textit{scheme} of types.
For example, the term \lstinline[language=Haskell]{let f = \x. x in (f 1, f True)}
 is not typeable in simply typed lambda calculus,
 since this requires \lstinline{f} to have both type $\textit{Int}\rightarrow\textit{Int}$
 and $\textit{Bool}\rightarrow\textit{Bool}$.
In contrast, A Hindley-Milner type system will accept this term,
 because both $\textit{Int}\rightarrow\textit{Int}$
 and $\textit{Bool}\rightarrow\textit{Bool}$ are instantiations of the type scheme
 $\forall \alpha,\alpha\rightarrow\alpha$.

Concretely,
 to infer a type for the above term,
 a type inference algorithm will first introduce a fresh type variable $\alpha$ for \lstinline{x},
 the argument to function \lstinline{f}, and infer that
 the type of \lstinline{f} is $\alpha\rightarrow\alpha$.
Next, because \lstinline{f} is bound in a let expression,
 the algorithm generalizes the type of \lstinline{f} to be a scheme by introducing forall quantified
 variables, i.e., $\forall \alpha.\alpha\rightarrow\alpha$.
At the call site of \lstinline{f},
 the type scheme is instantiated by consistently substituting forall quantified type variables with fresh ones,
 and the fresh type variables are later unified with concrete types.
For example, \lstinline{f} in function application \lstinline{f 1}
 may be instantiated with type $\alpha_1\rightarrow\alpha_1$.
Because integer \lstinline{1} has type \textit{Int},
 type variable $\alpha_1$ is unified with \textit{Int},
 making the occurrence of \lstinline{f} here
 have type $\textit{Int}\rightarrow\textit{Int}$.
The final type of \lstinline{f 1} is therefore \textit{Int}.

The key enabler of Hindley-Milner inference
 is the ability to unify two types.
To do this,
 an imperative implementation like Algorithm W \citep{hindley-milner} needs
 to track the alias graphs among type variables
 and potentially mutating a type variable
 to link to a concrete type,
 which requires careful handling.
In contrast, \egglog has the right abstractions for Hindley-Milner inference
 with the built-in power of unification.
The unification mechanism can be expressed as a single injectivity rule
\begin{lstlisting}[language=egglog]
(rule ((= (Arr fr1 to1) (Arr fr2 to2)))
      ((union fr1 fr2)
       (union to1 to2)))
\end{lstlisting}
This rule propagates unification down
 from inductively defined types to their children types.
At unification sites,
 it suffices to call \lstinline[language=egglog]{union}
 on the types being unified.
For instance, calling \lstinline[language=egglog]{union} on
\lstinline{(Arr (TVar x) (Int))} and \lstinline{(Arr (Bool) (TVar y))}
 will unify type variable $x$ (resp.\ $y$) and \textit{Int} (resp.\ \textit{Bool})
 by putting them into the same \eclass.

\begin{figure}
\begin{minipage}[t]{0.48\linewidth}
    \begin{lstlisting}[language=egglog]
(function generalize (Ctx Type) Scheme)
(function instantiate (Scheme) Type)
(function lookup (Ctx Ident) Scheme)
(function typeof (Ctx Expr i64) Type)

;; Injectivity of unification
(rule ((= (Arr fr1 to1) (Arr fr2 to2)))
      ((union fr1 fr2)
       (union to1 to2)))

;; Simple types
(rewrite (typeof ctx (Num x)) (Int))
(rewrite (typeof ctx (True)) (Bool))
(rewrite (typeof ctx (False)) (Bool))
(rewrite (typeof ctx (Var x))
         (instantiate (lookup ctx x)))

;; Inferring types for lambda abstractions
(rule ((= t (typeof ctx (Abs x e))))
      ((define fresh-tv (TVar (Fresh x)))
       (define scheme (Forall (empty) fresh-tv))
       (define new-ctx (Cons x scheme ctx))
       (define t1 (typeof new-ctx e))
       (union t (Arr fresh-tv t1))))
    \end{lstlisting}
\end{minipage}
\hfill
\begin{minipage}[t]{0.48\linewidth}
    \begin{lstlisting}[language=egglog]
;; Inferring types for function applcations
(rule ((= t (typeof ctx (App f e))))
	  ((define t1 (typeof ctx f))
	   (define t2 (typeof ctx e))
	   (union t1 (TArr t2 t))))
;; Inferring types for let expressions
(rule ((= t (typeof ctx (Let x e1 e2))))
      ((define t1 (typeof ctx e1))
       (define scheme (generalize ctx t1))
       (define new-ctx (Cons x scheme ctx))
       (define t2 (typeof new-ctx e2))
       (union t t2)))

;; Occurs check
(relation occurs-check (Ident Type))
(relation errors (Ident))
(rule ((= (TVar x) (Arr fr to)))
      ((occurs-check x fr)
       (occurs-check x to)))
(rule ((occurs-check x (Arr fr to)))
      ((occurs-check x fr)
       (occurs-check x to)))
(rule ((occurs-check x (TVar x)))
      ((errors x)
       (panic "occurs check failed")))
    \end{lstlisting}
\end{minipage}
\caption{Expressing Hindley-Milner inference in \egglog.
 \lstinline{Ident} is a datatype for identifiers
  that can be constructed by lifting a string or a counter (i.e., \lstinline{i64}).
 In the actual implementation,
  we additionally track a counter in the \lstinline{typeof} function
  to ensure the freshness of fresh variables,
  which we omit for brevity.
 Definitions of \lstinline{instantiate}, \lstinline{generalize},
 and \lstinline{lookup} are not shown as well.
 }
 \label{fig:hm}
\end{figure}

\autoref{fig:hm} shows a snippet of Hindley-Milner inference in \egglog.
We translate the typing rule to rewrite rules in \egglog straightforwardly.
The \egglog rule for lambda abstractions says,
 whenever we see a demand to check the type of \lstinline[language=Haskell]{\x.e}.
We create a fresh type scheme \lstinline{fresh-tv} with no variables quantified,
 binding it to \lstinline{x} in the context, and infer the type of body as \lstinline{t1}.
Finally, we unify the type of  \lstinline[language=Haskell]{\x.e} with
 \lstinline[mathescape=true]{fresh-tv $\rightarrow$ t1}.
For function application \lstinline{f e},
 we can compact the two rules
 for function application in simply typed lambda calculus into one rules
 thanks to the injectivity rule:
 we simply equate the type \lstinline{t1} of \lstinline{f} and the arrow type \lstinline{Arr t t2}
 for type \lstinline{t} of \lstinline{f e} and type \lstinline{t2} of \lstinline{e},
 and injectivity will handle the rest of unifications.
Finally, the rule for type inferring \lstinline[language=Haskell]{let x = e1 in e2}
 will first get and,
 generalize\footnote{
    Generalization, as well as instantiation for the rule for type inferring variables
    is a standard operation in type inference literature.
    They convert between types and type schemes based on contexts.
    We omit them from the presentation for brevity.
    To implement them, we also track the free type variables
    for each type
    in our implementation.
 } the type of \lstinline{e1} in the current context,
 bind variable \lstinline{x} to it
 and infer the type of \lstinline{e2} as \lstinline{t2}.
The type of \lstinline[language=Haskell]{let x = e1 in e2}
 is then unified with the type of \lstinline{t2}.

In Hindley-Milner type systems,
 a type variable may be accidentally unified
 with a type that contains it, which results in infinitary types
 like $\alpha\rightarrow\alpha\rightarrow\ldots$ and is usually
 not what users intend to do.
A Hindley-Milner inference algorithm will also
 do an ``occurs check'' before
 unifying a type variable with a type.
In \egglog, the occurs check can be done modularly,
 completely independent of the unification mechanism
 and the type inference algorithm.
In \autoref{fig:hm},
 we define an \lstinline{occurs-check} relation
 and match on cases where a type variable
 is unified with an inductive type like the arrow type
 and mark types that need to be occurs checked by
 populating them in the \lstinline{occurs-check} relation.
The occurs check fails when an \lstinline{occurs-check} demand
 is populated on an identifier and a type variable with the same identifier.
Our actual implementation also contains rules that
 check if two different base types are unified or
 a base type is unified with an arrow type,
 where it will throw an error.
These could happen when two incompatible types are unified
 due to ill-typed terms
 (e.g., when type inferring \lstinline[language=Haskell]{True + 1}).

\subsection{Other \egglog Pearls}\label{sec:other-pearls}

In this subsection,
 we show more self-contained programs with interesting behaviors,
 further demonstrating the expressive power of \egglog.

\paragraph{Equation Solving}
Many uses of \eqsat and hence \egglog fall into a guarded rewriting paradigm. A
different mode of use is that of equation solving: rather than
taking a left hand side and producing a right, \egglog can take an entire equation
and produces a new equation.
A common manipulation in algebraic reasoning is to
manipulate equations by applying the same operation to both sides. This is often
 used to isolate variables
 and use one variable to substitute other variables in an equation.
Substitutions in \egraphs and \egglog are implicit
 (since the variable and its definition via other variables are in the same equivalence class),
 and we can encode variable isolation as rules.

\autoref{fig:equationsolve:egglog} shows a simplistic equational system
 with addition, multiplication, and negations.
Besides the standard algebraic rules,
 we use two rules that manipulate equations to isolate variables.
This allows us to solve simple multivariable equations
 like $\begin{cases}
    z+y=6\\
    2z=y
 \end{cases}$.

Equation solving in \egglog can be seen as similar to the ``random walk'' approach to
 variable elimination a student may take. For specific solvable systems this may
 be very inefficient compared to a dedicated algorithm.
 For example one can
 consider a symbolic representation of a linear algebraic system,
 for which Gaussian elimination will be vastly more efficient.
However, equation solving in \egglog is
 compositional and
 can easily handle the addition of new domain-specific rules
 like those for trigonometric functions.

\begin{figure}
\begin{minipage}[t]{0.48\linewidth}
    \begin{lstlisting}[language=egglog]
(datatype Expr
    (Add Expr Expr)
    (Mul Expr Expr)
    (Neg Expr)
    (Num i64)
    (Var String))

;; Algebraic rules over expressions
(rewrite (Add x y) (Add y x))
(rewrite (Add (Add x y) z) (Add x (Add y z)))
(rewrite (Add (Mul y x) (Mul z x))
         (Mul (Add y z) x))

;; Make the implicit coefficient 1 explicit
(rewrite (Var x) (Mul (Num 1) (Var x)))

;; Constant folding
(rewrite (Add (Num x) (Num y)) (Num (+ x y)))
(rewrite (Neg (Num n)) (Num (- n)))
(rewrite (Add (Neg x) x) (Num 0))
    \end{lstlisting}
\end{minipage}
\hfill
\begin{minipage}[t]{0.48\linewidth}
    \begin{lstlisting}[language=egglog]
;; Variable isolation by rewriting
;; the entire equation
(rule ((= (Add x y) z))
      ((union (Add z (Neg y)) x)))
(rule ((= (Mul (Num x) y) (Num z))
       (= (% z x) 0))
      ((union (Num (/ z x)) y)))

; system 1: x + 2 = 7
(set (Add (Var "x") (Num 2)) (Num 7))
; system 2: z + y = 6; 2z = y
(set (Add (Var "z") (Var "y")) (Num 6))
(set (Add (Var "z") (Var "z")) (Var "y"))

(run 5) ;; run 5 iterations

(extract (Var "x")) ;; (Num 5)
(extract (Var "y")) ;; (Num 4)
(extract (Var "z")) ;; (Num 2)
    \end{lstlisting}
\end{minipage}
    \caption{Equation Solving in \egglog.}
    \label{fig:equationsolve:egglog}
\end{figure}

\begin{figure}
\begin{minipage}[t]{0.48\linewidth}
    \begin{lstlisting}[language=egglog]
;; Proofs of connectivity
(datatype Proof
    (Trans i64 Proof)
    (Edge i64 i64))

;; Path function points to a proof datatype
(function path (i64 i64) Proof)
(relation edge (i64 i64))

;; Base case
(rule ((edge x y))
      ((set (path x y) (Edge x y))))
\end{lstlisting}
\end{minipage}
\hfill
\begin{minipage}[t]{0.48\linewidth}
    \begin{lstlisting}[language=egglog]
;; Inductive case
(rule ((edge x y) (= p (path y z)))
      ((set (path x z) (Trans x p))))

;; Populate the graph and run
(edge 1 2)
(edge 2 3)
(edge 1 3)
(run)

;; returns the smallest proof of
;; the connectivity between 1 and 3
(extract (path 1 3))
    \end{lstlisting}
    \end{minipage}
    \caption{Encoding compact proofs in \egglog}
    \label{fig:proofs:egglog}
\end{figure}

\paragraph{Proof Datatypes}
Datalog proofs can be internalized as syntax trees inside of \egglog. This proof
datatype has one constructor for every Datalog rule of the program and records
any intermediate information that may be necessary. This can also be done in any
Datalog system that supports terms. A unique capability of \egglog however is the
ability to consider proofs of the same fact to be equivalent, a form of proof
irrelevance. This compresses the space used to store the proofs and enhances the
termination of the program which would not terminate in ordinary. In addition,
the standard extraction procedure can be used to extract a short proof.

\begin{figure}
\begin{minipage}[t]{0.42\linewidth}
    \begin{lstlisting}[language=egglog]
(datatype MExpr
    (MMul MExpr MExpr)
    (Kron MExpr MExpr)
    (Var String))

(datatype Dim
    (Times Dim Dim)
    (NamedDim String)
    (Lit i64))

(function nrows (MExpr) Dim)
(function ncols (MExpr) Dim)

;; Computing the dimensions of
;; matrix expressions
(rewrite (nrows (Kron A B))
         (Times (nrows A) (nrows B)))
(rewrite (ncols (Kron A B))
         (Times (ncols A) (ncols B)))
(rewrite (nrows (MMul A B)) (nrows A))
(rewrite (ncols (MMul A B)) (ncols B))
\end{lstlisting}
\end{minipage}
\begin{minipage}[t]{0.50\linewidth}
    \begin{lstlisting}[language=egglog]
;; Reasoning about dimensionality
(rewrite (Times a (Times b c))
         (Times (Times a b) c))
(rewrite (Times (Lit i) (Lit j)) (Lit (* i j)))
(rewrite (Times a b) (Times b a))

;; Rewriting matrix multiplications and Kronecker
;; product
(rewrite (MMul A (MMul B C)) (MMul (MMul A B) C))
(rewrite (MMul (MMul A B) C) (MMul A (MMul B C)))
(rewrite (Kron A (Kron B C)) (Kron (Kron A B) C))
(rewrite (Kron (Kron A B) C) (Kron A (Kron B C)))
(rewrite (Kron (MMul A C) (MMul B D))
         (MMul (Kron A B) (Kron C D)))

;; Optimizing Kronecker product with guarded rules
(rewrite (MMul (Kron A B) (Kron C D))
         (Kron (MMul A C) (MMul B D))
    :when ((= (ncols A) (nrows C))
           (= (ncols B) (nrows D))))
\end{lstlisting}
\end{minipage}
\caption{Equality saturation with matrices in \egglog.
The last rule is guarded by the equational precondition
 that the dimensionalities should align,
 which is made possible by
 rich semantic analyses \textit{a la} Datalog.}
\label{fig:matrix:egglog}
\end{figure}

\paragraph{Reasoning about matrices}
The algebra of matrices follows similar rules as the algebra of simple numbers,
 except that matrix multiplication generally not commutative.
With the addition of structural operations like the Kronecker product, direct sum, and stacking of matrices a richer algebraic structure emerges.
A particularly simple and useful rewrite rule allows one to
 push matrix multiplication through a Kronecker product
 $(A \otimes B) \cdot (C \otimes D) = (A \cdot C) \otimes (B \cdot D)$.
Rewriting from left to right improves the asymptotic complexity of evaluating the expression.
However,
 while this equation may proceed from right to left unconditionally,
 the left to right application requires that the dimensionality of the matrices line up.
In a large matrix expression with possibly abstract dimensionality,
 this is not easily doable in a classical equality saturation framework.
Although one may be tempted to express dimensionality
 with \eclass analyses,
 the dimensionality is a symbolic term itself and needs to be reasoned about via algebraic rewriting.
However, the abstraction of \eclass analyses do not allow rewriting over the analysis values.
On the other hand, because the analysis is just another function
 not unlike the constructors for matrix expressions,
 we can use standard \egglog rules to reason about it
 just like how we reason about matrix expressions.
\autoref{fig:matrix:egglog} shows
 a simple theory of matrices with Kronecker product,
 and this example can be generalized to
 other (essentially) algebraic theories.

%% file: sn.tex
\section{Correctness of the \seminaive algorithm}\label{sec:sn-proof}

In this section, we show the correctness of the \seminaive algorithm.

\begin{theorem}
    The \seminaive evaluation of an \egglog program
    produces the same database as the \naive
    evaluation.
\end{theorem}
\begin{proof}
    We use $I^\SN_i=(\DB^\SN_i, \equiv^\SN_i)$ and $I^\N_i=(\DB^\N_i, \equiv^\N_i)$ 
    to denote the instances 
    produced by the \seminaive evaluation
    and \naive evaluation.   
    We prove $I^\SN_i=I^\N_i$ by induction. 
    It is easy to see $I^\SN_i=I^\N_i$ for $i=0, 1$,
    and we will prove that, for $i\geq 1$,
    if $I^\SN_j=I^\N_j$ for $j\leq i$, $I^\SN_{i+1}=I^\N_{i+1}$ holds.

    First, because $\Delta\DB{}_i = \DB{}^\SN_{i} - \DB{}^\SN_{i-1} = \DB{}^\N_{i} - \DB{}^\N_{i-1}$ by definition,
    we have \begin{align}
        \DB^\N_{i-1}&\supseteq \DB^\N_i - \Delta\DB_i,\\
        T_P(I^\N_{i-1})&\supseteq T_P(I^\N_i - \Delta\DB_i)\text{\quad by monotonicity of $T_P$ w.r.t.\ $\subseteq$},\\
        T_P(I^\N_{i-1})\cup T_P^{\SN}(I^\N_i, \Delta\DB_i)
        &\supseteq T_P(I^\N_i - \Delta\DB_i)\cup T_P^{\SN}(I^\SN_i, \Delta\DB_i), \\
        &\supseteq T_P(I^\N_i) \text{\quad by inspecting the definition of $T^\SN_P$.}\label{eqn:l1}
    \end{align}

    Second, it is straightforward to see that, for all instance $I$ and database $\DB$
    \begin{align}
        R^\infty (R^\infty(I)\cup \DB) = R^\infty (I\cup \DB) \label{eqn:l2}.
    \end{align}

    Combining these two, we get
    \begin{align}
        I_{i+1}^\SN &= R^\infty\left(I_i^{\SN} \cup T_P^{\SN} (I^\SN_i, \Delta\DB_i)\right),\\
        &= R^\infty\left(
            \underline{R^\infty\left(I^\N_{i-1}\cup T_P(I^\N_{i-1}) \right)}
        \cup T_P^\SN (I^\SN_i, \Delta\DB_i)\right)\text{\quad by the ind. hypothesis and defn. of $I^N_i$,}\\
        &= R^\infty\left(
            \underline{I^\N_{i-1}\cup T_P(I^N_{i-1})}
        \cup T_P^\SN (I^\SN_i, \Delta\DB_i)\right)\text{\quad by Eqn.~\ref{eqn:l2},}\\
        &= R^\infty\left(I^\N_{i-1}\cup T_P(I^N_{i-1}) 
        \cup T_P^\SN (I^\SN_i, \Delta\DB_i)
        \underline{
            \,\cup T_P(I^\N_i)
        }\right)\text{\quad 
        by Eqn.~\ref{eqn:l1},
        }\\
        &= R^\infty\left(I^\N_{i-1}\cup T_P(I^N_{i-1}) 
        \cup T_P(I^\N_i)\right)\text{\quad 
        since $T_P^\SN (I^\SN_i, \Delta\DB_i)$ is a subset of $T_P(I^\N_i)$,}\\
        &=R^\infty\left(
            \underline{
                R^\infty\left(I^\N_{i-1}\cup T_P(I^N_{i-1})\right) 
            }
        \cup T_P(I^\N_i)\right)\text{\quad
        by Eqn.~\ref{eqn:l2},}\\
        &=R^\infty\left(\underline{I^\N_i}
        \cup T_P(I^\N_i)\right)=I_{i+1}^\N.
    \end{align}

\end{proof}

%% file: paper.bbl

\begin{thebibliography}{65}


\ifx \showCODEN    \undefined \def \showCODEN     #1{\unskip}     \fi
\ifx \showDOI      \undefined \def \showDOI       #1{#1}\fi
\ifx \showISBNx    \undefined \def \showISBNx     #1{\unskip}     \fi
\ifx \showISBNxiii \undefined \def \showISBNxiii  #1{\unskip}     \fi
\ifx \showISSN     \undefined \def \showISSN      #1{\unskip}     \fi
\ifx \showLCCN     \undefined \def \showLCCN      #1{\unskip}     \fi
\ifx \shownote     \undefined \def \shownote      #1{#1}          \fi
\ifx \showarticletitle \undefined \def \showarticletitle #1{#1}   \fi
\ifx \showURL      \undefined \def \showURL       {\relax}        \fi
\providecommand\bibfield[2]{#2}
\providecommand\bibinfo[2]{#2}
\providecommand\natexlab[1]{#1}
\providecommand\showeprint[2][]{arXiv:#2}

\bibitem[\protect\citeauthoryear{Abiteboul, Hull, and Vianu}{Abiteboul
  et~al\mbox{.}}{1995}]%
        {alice-book}
\bibfield{author}{\bibinfo{person}{Serge Abiteboul}, \bibinfo{person}{Richard
  Hull}, {and} \bibinfo{person}{Victor Vianu}.}
  \bibinfo{year}{1995}\natexlab{}.
\newblock \bibinfo{booktitle}{\emph{Foundations of Databases}}.
\newblock \bibinfo{publisher}{Addison-Wesley}.
\newblock
\showISBNx{0-201-53771-0}
\urldef\tempurl%
\url{http://webdam.inria.fr/Alice/}
\showURL{%
\tempurl}


\bibitem[\protect\citeauthoryear{Abo~Khamis, Ngo, Pichler, Suciu, and
  Wang}{Abo~Khamis et~al\mbox{.}}{2022}]%
        {datalogo}
\bibfield{author}{\bibinfo{person}{Mahmoud Abo~Khamis},
  \bibinfo{person}{Hung~Q. Ngo}, \bibinfo{person}{Reinhard Pichler},
  \bibinfo{person}{Dan Suciu}, {and} \bibinfo{person}{Yisu~Remy Wang}.}
  \bibinfo{year}{2022}\natexlab{}.
\newblock \showarticletitle{Convergence of Datalog over (Pre-) Semirings}. In
  \bibinfo{booktitle}{\emph{Proceedings of the 41st ACM SIGMOD-SIGACT-SIGAI
  Symposium on Principles of Database Systems}} (Philadelphia, PA, USA)
  \emph{(\bibinfo{series}{PODS '22})}. \bibinfo{publisher}{Association for
  Computing Machinery}, \bibinfo{address}{New York, NY, USA},
  \bibinfo{pages}{105–117}.
\newblock
\showISBNx{9781450392600}
\urldef\tempurl%
\url{https://doi.org/10.1145/3517804.3524140}
\showDOI{\tempurl}


\bibitem[\protect\citeauthoryear{Aref, ten Cate, Green, Kimelfeld, Olteanu,
  Pasalic, Veldhuizen, and Washburn}{Aref et~al\mbox{.}}{2015}]%
        {logicblox}
\bibfield{author}{\bibinfo{person}{Molham Aref}, \bibinfo{person}{Balder ten
  Cate}, \bibinfo{person}{Todd~J. Green}, \bibinfo{person}{Benny Kimelfeld},
  \bibinfo{person}{Dan Olteanu}, \bibinfo{person}{Emir Pasalic},
  \bibinfo{person}{Todd~L. Veldhuizen}, {and} \bibinfo{person}{Geoffrey
  Washburn}.} \bibinfo{year}{2015}\natexlab{}.
\newblock \showarticletitle{Design and Implementation of the LogicBlox System}.
  In \bibinfo{booktitle}{\emph{Proceedings of the 2015 ACM SIGMOD International
  Conference on Management of Data}} (Melbourne, Victoria, Australia)
  \emph{(\bibinfo{series}{SIGMOD '15})}. \bibinfo{publisher}{Association for
  Computing Machinery}, \bibinfo{address}{New York, NY, USA},
  \bibinfo{pages}{1371–1382}.
\newblock
\showISBNx{9781450327589}
\urldef\tempurl%
\url{https://doi.org/10.1145/2723372.2742796}
\showDOI{\tempurl}


\bibitem[\protect\citeauthoryear{Balatsouras and Smaragdakis}{Balatsouras and
  Smaragdakis}{2016}]%
        {cclyzer}
\bibfield{author}{\bibinfo{person}{George Balatsouras} {and}
  \bibinfo{person}{Yannis Smaragdakis}.} \bibinfo{year}{2016}\natexlab{}.
\newblock \showarticletitle{Structure-Sensitive Points-To Analysis for {C} and
  {C++}}. In \bibinfo{booktitle}{\emph{Static Analysis - 23rd International
  Symposium, {SAS} 2016, Edinburgh, UK, September 8-10, 2016, Proceedings}}
  \emph{(\bibinfo{series}{Lecture Notes in Computer Science},
  Vol.~\bibinfo{volume}{9837})}, \bibfield{editor}{\bibinfo{person}{Xavier
  Rival}} (Ed.). \bibinfo{publisher}{Springer}, \bibinfo{pages}{84--104}.
\newblock
\urldef\tempurl%
\url{https://doi.org/10.1007/978-3-662-53413-7\_5}
\showDOI{\tempurl}


\bibitem[\protect\citeauthoryear{Balbin and Ramamohanarao}{Balbin and
  Ramamohanarao}{1987}]%
        {seminaive}
\bibfield{author}{\bibinfo{person}{Isaac Balbin} {and}
  \bibinfo{person}{Kotagiri Ramamohanarao}.} \bibinfo{year}{1987}\natexlab{}.
\newblock \showarticletitle{A Generalization of the Differential Approach to
  Recursive Query Evaluation}.
\newblock \bibinfo{journal}{\emph{J. Log. Program.}} \bibinfo{volume}{4},
  \bibinfo{number}{3} (\bibinfo{date}{sep} \bibinfo{year}{1987}),
  \bibinfo{pages}{259–262}.
\newblock
\showISSN{0743-1066}
\urldef\tempurl%
\url{https://doi.org/10.1016/0743-1066(87)90004-5}
\showDOI{\tempurl}


\bibitem[\protect\citeauthoryear{Barrett, Conway, Deters, Hadarean,
  Jovanovi\'{c}, King, Reynolds, and Tinelli}{Barrett et~al\mbox{.}}{2011}]%
        {cvc4}
\bibfield{author}{\bibinfo{person}{Clark Barrett},
  \bibinfo{person}{Christopher~L. Conway}, \bibinfo{person}{Morgan Deters},
  \bibinfo{person}{Liana Hadarean}, \bibinfo{person}{Dejan Jovanovi\'{c}},
  \bibinfo{person}{Tim King}, \bibinfo{person}{Andrew Reynolds}, {and}
  \bibinfo{person}{Cesare Tinelli}.} \bibinfo{year}{2011}\natexlab{}.
\newblock \showarticletitle{CVC4}. In \bibinfo{booktitle}{\emph{Proceedings of
  the 23rd International Conference on Computer Aided Verification}} (Snowbird,
  UT) \emph{(\bibinfo{series}{CAV'11})}. \bibinfo{publisher}{Springer-Verlag},
  \bibinfo{address}{Berlin, Heidelberg}, \bibinfo{pages}{171–177}.
\newblock
\showISBNx{9783642221095}


\bibitem[\protect\citeauthoryear{Barrett and Moore}{Barrett and Moore}{2022}]%
        {cclyzerpp}
\bibfield{author}{\bibinfo{person}{Langston Barrett} {and}
  \bibinfo{person}{Scott Moore}.} \bibinfo{year}{2022}\natexlab{}.
\newblock \bibinfo{title}{cclyzer++: Scalable and Precise Pointer Analysis for
  LLVM}.
\newblock
  \bibinfo{howpublished}{\url{https://galois.com/blog/2022/08/cclyzer-scalable-and-precise-pointer-analysis-for-llvm/}}.
\newblock


\bibitem[\protect\citeauthoryear{Bellomarini, Sallinger, and
  Gottlob}{Bellomarini et~al\mbox{.}}{2018}]%
        {vadalog}
\bibfield{author}{\bibinfo{person}{Luigi Bellomarini}, \bibinfo{person}{Emanuel
  Sallinger}, {and} \bibinfo{person}{Georg Gottlob}.}
  \bibinfo{year}{2018}\natexlab{}.
\newblock \showarticletitle{The Vadalog System: Datalog-Based Reasoning for
  Knowledge Graphs}.
\newblock \bibinfo{journal}{\emph{Proc. VLDB Endow.}} \bibinfo{volume}{11},
  \bibinfo{number}{9} (\bibinfo{date}{may} \bibinfo{year}{2018}),
  \bibinfo{pages}{975–987}.
\newblock
\showISSN{2150-8097}
\urldef\tempurl%
\url{https://doi.org/10.14778/3213880.3213888}
\showDOI{\tempurl}


\bibitem[\protect\citeauthoryear{Bembenek, Greenberg, and Chong}{Bembenek
  et~al\mbox{.}}{2020}]%
        {DBLP:journals/pacmpl/Bembenek0C20}
\bibfield{author}{\bibinfo{person}{Aaron Bembenek}, \bibinfo{person}{Michael
  Greenberg}, {and} \bibinfo{person}{Stephen Chong}.}
  \bibinfo{year}{2020}\natexlab{}.
\newblock \showarticletitle{Formulog: Datalog for SMT-based static analysis}.
\newblock \bibinfo{journal}{\emph{Proc. {ACM} Program. Lang.}}
  \bibinfo{volume}{4}, \bibinfo{number}{{OOPSLA}} (\bibinfo{year}{2020}),
  \bibinfo{pages}{141:1--141:31}.
\newblock
\urldef\tempurl%
\url{https://doi.org/10.1145/3428209}
\showDOI{\tempurl}


\bibitem[\protect\citeauthoryear{Benedikt, Konstantinidis, Mecca, Motik,
  Papotti, Santoro, and Tsamoura}{Benedikt et~al\mbox{.}}{2017}]%
        {bench-chase}
\bibfield{author}{\bibinfo{person}{Michael Benedikt}, \bibinfo{person}{George
  Konstantinidis}, \bibinfo{person}{Giansalvatore Mecca},
  \bibinfo{person}{Boris Motik}, \bibinfo{person}{Paolo Papotti},
  \bibinfo{person}{Donatello Santoro}, {and} \bibinfo{person}{Efthymia
  Tsamoura}.} \bibinfo{year}{2017}\natexlab{}.
\newblock \showarticletitle{Benchmarking the Chase}. In
  \bibinfo{booktitle}{\emph{Proceedings of the 36th ACM SIGMOD-SIGACT-SIGAI
  Symposium on Principles of Database Systems}} (Chicago, Illinois, USA)
  \emph{(\bibinfo{series}{PODS '17})}. \bibinfo{publisher}{Association for
  Computing Machinery}, \bibinfo{address}{New York, NY, USA},
  \bibinfo{pages}{37–52}.
\newblock
\showISBNx{9781450341981}
\urldef\tempurl%
\url{https://doi.org/10.1145/3034786.3034796}
\showDOI{\tempurl}


\bibitem[\protect\citeauthoryear{Bidlingmaier}{Bidlingmaier}{2023a}]%
        {bidlingmaier2023algebraic}
\bibfield{author}{\bibinfo{person}{Martin~E. Bidlingmaier}.}
  \bibinfo{year}{2023}\natexlab{a}.
\newblock \bibinfo{title}{Algebraic Semantics of Datalog with Equality}.
\newblock
\newblock
\showeprint[arxiv]{2302.03167}~[cs.LO]


\bibitem[\protect\citeauthoryear{Bidlingmaier}{Bidlingmaier}{2023b}]%
        {bidlingmaier2023evaluation}
\bibfield{author}{\bibinfo{person}{Martin~E. Bidlingmaier}.}
  \bibinfo{year}{2023}\natexlab{b}.
\newblock \bibinfo{title}{An Evaluation Algorithm for Datalog with Equality}.
\newblock
\newblock
\showeprint[arxiv]{2302.05792}~[cs.PL]


\bibitem[\protect\citeauthoryear{Bravenboer and Smaragdakis}{Bravenboer and
  Smaragdakis}{2009}]%
        {doop}
\bibfield{author}{\bibinfo{person}{Martin Bravenboer} {and}
  \bibinfo{person}{Yannis Smaragdakis}.} \bibinfo{year}{2009}\natexlab{}.
\newblock \showarticletitle{Strictly Declarative Specification of Sophisticated
  Points-to Analyses}. In \bibinfo{booktitle}{\emph{Proceedings of the 24th ACM
  SIGPLAN Conference on Object Oriented Programming Systems Languages and
  Applications}} (Orlando, Florida, USA) \emph{(\bibinfo{series}{OOPSLA '09})}.
  \bibinfo{publisher}{Association for Computing Machinery},
  \bibinfo{address}{New York, NY, USA}, \bibinfo{pages}{243–262}.
\newblock
\showISBNx{9781605587660}
\urldef\tempurl%
\url{https://doi.org/10.1145/1640089.1640108}
\showDOI{\tempurl}


\bibitem[\protect\citeauthoryear{Cal\`{\i}, Gottlob, and Lukasiewicz}{Cal\`{\i}
  et~al\mbox{.}}{2009}]%
        {datalogpm}
\bibfield{author}{\bibinfo{person}{Andrea Cal\`{\i}}, \bibinfo{person}{Georg
  Gottlob}, {and} \bibinfo{person}{Thomas Lukasiewicz}.}
  \bibinfo{year}{2009}\natexlab{}.
\newblock \showarticletitle{A General Datalog-Based Framework for Tractable
  Query Answering over Ontologies}. In \bibinfo{booktitle}{\emph{Proceedings of
  the Twenty-Eighth ACM SIGMOD-SIGACT-SIGART Symposium on Principles of
  Database Systems}} (Providence, Rhode Island, USA)
  \emph{(\bibinfo{series}{PODS '09})}. \bibinfo{publisher}{Association for
  Computing Machinery}, \bibinfo{address}{New York, NY, USA},
  \bibinfo{pages}{77–86}.
\newblock
\showISBNx{9781605585536}
\urldef\tempurl%
\url{https://doi.org/10.1145/1559795.1559809}
\showDOI{\tempurl}


\bibitem[\protect\citeauthoryear{Cheli}{Cheli}{2021}]%
        {metatheory.jl}
\bibfield{author}{\bibinfo{person}{Alessandro Cheli}.}
  \bibinfo{year}{2021}\natexlab{}.
\newblock \showarticletitle{Metatheory.jl: Fast and Elegant Algebraic
  Computation in Julia with Extensible Equality Saturation}.
\newblock \bibinfo{journal}{\emph{Journal of Open Source Software}}
  \bibinfo{volume}{6}, \bibinfo{number}{59} (\bibinfo{year}{2021}),
  \bibinfo{pages}{3078}.
\newblock
\urldef\tempurl%
\url{https://doi.org/10.21105/joss.03078}
\showDOI{\tempurl}


\bibitem[\protect\citeauthoryear{Conway, Marczak, Alvaro, Hellerstein, and
  Maier}{Conway et~al\mbox{.}}{2012}]%
        {bloom-lattice}
\bibfield{author}{\bibinfo{person}{Neil Conway}, \bibinfo{person}{William~R.
  Marczak}, \bibinfo{person}{Peter Alvaro}, \bibinfo{person}{Joseph~M.
  Hellerstein}, {and} \bibinfo{person}{David Maier}.}
  \bibinfo{year}{2012}\natexlab{}.
\newblock \showarticletitle{Logic and lattices for distributed programming}. In
  \bibinfo{booktitle}{\emph{{ACM} Symposium on Cloud Computing, {SOCC} '12, San
  Jose, CA, USA, October 14-17, 2012}}. \bibinfo{pages}{1}.
\newblock
\urldef\tempurl%
\url{https://doi.org/10.1145/2391229.2391230}
\showDOI{\tempurl}


\bibitem[\protect\citeauthoryear{de~Moura and Bj{\o}rner}{de~Moura and
  Bj{\o}rner}{2007}]%
        {ematching}
\bibfield{author}{\bibinfo{person}{Leonardo de Moura} {and}
  \bibinfo{person}{Nikolaj Bj{\o}rner}.} \bibinfo{year}{2007}\natexlab{}.
\newblock \showarticletitle{Efficient E-Matching for SMT Solvers}. In
  \bibinfo{booktitle}{\emph{Automated Deduction -- CADE-21}},
  \bibfield{editor}{\bibinfo{person}{Frank Pfenning}} (Ed.).
  \bibinfo{publisher}{Springer Berlin Heidelberg}, \bibinfo{address}{Berlin,
  Heidelberg}, \bibinfo{pages}{183--198}.
\newblock
\showISBNx{978-3-540-73595-3}


\bibitem[\protect\citeauthoryear{De~Moura and Bj{\o}rner}{De~Moura and
  Bj{\o}rner}{2008}]%
        {z3}
\bibfield{author}{\bibinfo{person}{Leonardo De~Moura} {and}
  \bibinfo{person}{Nikolaj Bj{\o}rner}.} \bibinfo{year}{2008}\natexlab{}.
\newblock \showarticletitle{Z3: An Efficient SMT Solver}. In
  \bibinfo{booktitle}{\emph{Proceedings of the Theory and Practice of Software,
  14th International Conference on Tools and Algorithms for the Construction
  and Analysis of Systems}} (Budapest, Hungary)
  \emph{(\bibinfo{series}{TACAS'08/ETAPS'08})}.
  \bibinfo{publisher}{Springer-Verlag}, \bibinfo{address}{Berlin, Heidelberg},
  \bibinfo{pages}{337--340}.
\newblock
\showISBNx{3-540-78799-2, 978-3-540-78799-0}
\urldef\tempurl%
\url{http://dl.acm.org/citation.cfm?id=1792734.1792766}
\showURL{%
\tempurl}


\bibitem[\protect\citeauthoryear{Detlefs, Nelson, and Saxe}{Detlefs
  et~al\mbox{.}}{2005}]%
        {simplify}
\bibfield{author}{\bibinfo{person}{David Detlefs}, \bibinfo{person}{Greg
  Nelson}, {and} \bibinfo{person}{James~B. Saxe}.}
  \bibinfo{year}{2005}\natexlab{}.
\newblock \showarticletitle{Simplify: A Theorem Prover for Program Checking}.
\newblock \bibinfo{journal}{\emph{J. ACM}} \bibinfo{volume}{52},
  \bibinfo{number}{3} (\bibinfo{date}{May} \bibinfo{year}{2005}),
  \bibinfo{pages}{365--473}.
\newblock
\showISSN{0004-5411}
\urldef\tempurl%
\url{https://doi.org/10.1145/1066100.1066102}
\showDOI{\tempurl}


\bibitem[\protect\citeauthoryear{Deutsch, Nash, and Remmel}{Deutsch
  et~al\mbox{.}}{2008}]%
        {chase-revisited}
\bibfield{author}{\bibinfo{person}{Alin Deutsch}, \bibinfo{person}{Alan Nash},
  {and} \bibinfo{person}{Jeff Remmel}.} \bibinfo{year}{2008}\natexlab{}.
\newblock \showarticletitle{The Chase Revisited}. In
  \bibinfo{booktitle}{\emph{Proceedings of the Twenty-Seventh ACM
  SIGMOD-SIGACT-SIGART Symposium on Principles of Database Systems}}
  (Vancouver, Canada) \emph{(\bibinfo{series}{PODS '08})}.
  \bibinfo{publisher}{Association for Computing Machinery},
  \bibinfo{address}{New York, NY, USA}, \bibinfo{pages}{149–158}.
\newblock
\showISBNx{9781605581521}
\urldef\tempurl%
\url{https://doi.org/10.1145/1376916.1376938}
\showDOI{\tempurl}


\bibitem[\protect\citeauthoryear{developers}{developers}{[n.d.]}]%
        {rel-refernce}
\bibfield{author}{\bibinfo{person}{Rel developers}.}
  \bibinfo{year}{[n.d.]}\natexlab{}.
\newblock \bibinfo{title}{Rel reference}.
\newblock
\newblock
\urldef\tempurl%
\url{https://docs.relational.ai/rel/ref/overview}
\showURL{%
\tempurl}


\bibitem[\protect\citeauthoryear{Developers}{Developers}{[n.d.]}]%
        {souffle-adt}
\bibfield{author}{\bibinfo{person}{Souffl{\'e} Developers}.}
  \bibinfo{year}{[n.d.]}\natexlab{}.
\newblock \bibinfo{title}{Souffl{\'e} Algebraic Data Types}.
\newblock
  \bibinfo{howpublished}{\url{https://souffle-lang.github.io/types\#algebraic-data-types-adt}}.
\newblock
\newblock
\shownote{Accessed: 2022-11-01.}


\bibitem[\protect\citeauthoryear{Downey, Sethi, and Tarjan}{Downey
  et~al\mbox{.}}{1980}]%
        {tarjan-congruence}
\bibfield{author}{\bibinfo{person}{{Peter J.} Downey}, \bibinfo{person}{Ravi
  Sethi}, {and} \bibinfo{person}{{Robert Endre} Tarjan}.}
  \bibinfo{year}{1980}\natexlab{}.
\newblock \showarticletitle{Variations on the Common Subexpression Problem}.
\newblock \bibinfo{journal}{\emph{J. ACM}} \bibinfo{volume}{27},
  \bibinfo{number}{4} (\bibinfo{date}{1 Oct.} \bibinfo{year}{1980}),
  \bibinfo{pages}{758--771}.
\newblock
\showISSN{0004-5411}
\urldef\tempurl%
\url{https://doi.org/10.1145/322217.322228}
\showDOI{\tempurl}


\bibitem[\protect\citeauthoryear{Fagin, Kolaitis, Miller, and Popa}{Fagin
  et~al\mbox{.}}{2003}]%
        {data-exchange}
\bibfield{author}{\bibinfo{person}{Ronald Fagin}, \bibinfo{person}{Phokion~G.
  Kolaitis}, \bibinfo{person}{Ren{\'e}e~J. Miller}, {and}
  \bibinfo{person}{Lucian Popa}.} \bibinfo{year}{2003}\natexlab{}.
\newblock \showarticletitle{Data Exchange: Semantics and Query Answering}. In
  \bibinfo{booktitle}{\emph{Database Theory --- ICDT 2003}},
  \bibfield{editor}{\bibinfo{person}{Diego Calvanese},
  \bibinfo{person}{Maurizio Lenzerini}, {and} \bibinfo{person}{Rajeev Motwani}}
  (Eds.). \bibinfo{publisher}{Springer Berlin Heidelberg},
  \bibinfo{address}{Berlin, Heidelberg}, \bibinfo{pages}{207--224}.
\newblock
\showISBNx{978-3-540-36285-2}


\bibitem[\protect\citeauthoryear{Flatt, Coward, Willsey, Tatlock, and
  Panchekha}{Flatt et~al\mbox{.}}{2022}]%
        {flatt2022small}
\bibfield{author}{\bibinfo{person}{Oliver Flatt}, \bibinfo{person}{Samuel
  Coward}, \bibinfo{person}{Max Willsey}, \bibinfo{person}{Zachary Tatlock},
  {and} \bibinfo{person}{Pavel Panchekha}.} \bibinfo{year}{2022}\natexlab{}.
\newblock \showarticletitle{Small Proofs from Congruence Closure}. In
  \bibinfo{booktitle}{\emph{Proceedings of The 22nd Conference on Formal
  Methods in Computer-Aided Design (FMCAD '22)}}, Vol.~\bibinfo{volume}{3}. TU
  Wien Academic Press, \bibinfo{pages}{75}.
\newblock
\urldef\tempurl%
\url{https://doi.org/10.34727/2022/isbn.978-3-85448-053-2_13}
\showURL{%
\tempurl}


\bibitem[\protect\citeauthoryear{Frühwirth}{Frühwirth}{1998}]%
        {chr}
\bibfield{author}{\bibinfo{person}{Thom Frühwirth}.}
  \bibinfo{year}{1998}\natexlab{}.
\newblock \showarticletitle{Theory and practice of constraint handling rules}.
\newblock \bibinfo{journal}{\emph{The Journal of Logic Programming}}
  \bibinfo{volume}{37}, \bibinfo{number}{1} (\bibinfo{year}{1998}),
  \bibinfo{pages}{95--138}.
\newblock
\showISSN{0743-1066}
\urldef\tempurl%
\url{https://doi.org/10.1016/S0743-1066(98)10005-5}
\showDOI{\tempurl}


\bibitem[\protect\citeauthoryear{Hu, Karp, Zhao, Zreika, Wu, and Scholz}{Hu
  et~al\mbox{.}}{2021}]%
        {choice-souffle}
\bibfield{author}{\bibinfo{person}{Xiaowen Hu}, \bibinfo{person}{Joshua Karp},
  \bibinfo{person}{David Zhao}, \bibinfo{person}{Abdul Zreika},
  \bibinfo{person}{Xi Wu}, {and} \bibinfo{person}{Bernhard Scholz}.}
  \bibinfo{year}{2021}\natexlab{}.
\newblock \showarticletitle{The Choice Construct in the Souffl\'{e} Language}.
  In \bibinfo{booktitle}{\emph{Programming Languages and Systems: 19th Asian
  Symposium, APLAS 2021, Chicago, IL, USA, October 17–18, 2021, Proceedings}}
  (Chicago, IL, USA). \bibinfo{publisher}{Springer-Verlag},
  \bibinfo{address}{Berlin, Heidelberg}, \bibinfo{pages}{163–181}.
\newblock
\showISBNx{978-3-030-89050-6}
\urldef\tempurl%
\url{https://doi.org/10.1007/978-3-030-89051-3_10}
\showDOI{\tempurl}


\bibitem[\protect\citeauthoryear{Jordan, Scholz, and Suboti{\'c}}{Jordan
  et~al\mbox{.}}{2016}]%
        {souffle}
\bibfield{author}{\bibinfo{person}{Herbert Jordan}, \bibinfo{person}{Bernhard
  Scholz}, {and} \bibinfo{person}{Pavle Suboti{\'c}}.}
  \bibinfo{year}{2016}\natexlab{}.
\newblock \showarticletitle{Souffl{\'e}: On synthesis of program analyzers}. In
  \bibinfo{booktitle}{\emph{International Conference on Computer Aided
  Verification}}. Springer, \bibinfo{pages}{422--430}.
\newblock


\bibitem[\protect\citeauthoryear{Joshi, Nelson, and Randall}{Joshi
  et~al\mbox{.}}{2002}]%
        {denali}
\bibfield{author}{\bibinfo{person}{Rajeev Joshi}, \bibinfo{person}{Greg
  Nelson}, {and} \bibinfo{person}{Keith Randall}.}
  \bibinfo{year}{2002}\natexlab{}.
\newblock \showarticletitle{Denali: A Goal-directed Superoptimizer}.
\newblock \bibinfo{journal}{\emph{SIGPLAN Not.}} \bibinfo{volume}{37},
  \bibinfo{number}{5} (\bibinfo{date}{May} \bibinfo{year}{2002}),
  \bibinfo{pages}{304--314}.
\newblock
\showISSN{0362-1340}
\urldef\tempurl%
\url{https://doi.org/10.1145/543552.512566}
\showDOI{\tempurl}


\bibitem[\protect\citeauthoryear{Kanellakis and Revesz}{Kanellakis and
  Revesz}{1989}]%
        {congr-duality}
\bibfield{author}{\bibinfo{person}{Paris~C. Kanellakis} {and}
  \bibinfo{person}{Peter~Z. Revesz}.} \bibinfo{year}{1989}\natexlab{}.
\newblock \showarticletitle{On the relationship of congruence closureand
  unification}.
\newblock \bibinfo{journal}{\emph{Journal of Symbolic Computation}}
  \bibinfo{volume}{7}, \bibinfo{number}{3} (\bibinfo{year}{1989}),
  \bibinfo{pages}{427--444}.
\newblock
\showISSN{0747-7171}
\urldef\tempurl%
\url{https://doi.org/10.1016/S0747-7171(89)80018-5}
\showDOI{\tempurl}
\newblock
\shownote{Unification: Part 1.}


\bibitem[\protect\citeauthoryear{Koehler, Trinder, and Steuwer}{Koehler
  et~al\mbox{.}}{2021}]%
        {koehler2021sketch}
\bibfield{author}{\bibinfo{person}{Thomas Koehler}, \bibinfo{person}{Phil
  Trinder}, {and} \bibinfo{person}{Michel Steuwer}.}
  \bibinfo{year}{2021}\natexlab{}.
\newblock \showarticletitle{Sketch-Guided Equality Saturation: Scaling Equality
  Saturation to Complex Optimizations in Languages with Bindings}.
\newblock \bibinfo{journal}{\emph{arXiv preprint arXiv:2111.13040}}
  (\bibinfo{year}{2021}).
\newblock


\bibitem[\protect\citeauthoryear{Kolaitis and Papadimitriou}{Kolaitis and
  Papadimitriou}{1988}]%
        {neg-by-fixpoint}
\bibfield{author}{\bibinfo{person}{Phokion~G. Kolaitis} {and}
  \bibinfo{person}{Christos~H. Papadimitriou}.}
  \bibinfo{year}{1988}\natexlab{}.
\newblock \showarticletitle{Why Not Negation by Fixpoint?}. In
  \bibinfo{booktitle}{\emph{Proceedings of the Seventh ACM SIGACT-SIGMOD-SIGART
  Symposium on Principles of Database Systems}} (Austin, Texas, USA)
  \emph{(\bibinfo{series}{PODS '88})}. \bibinfo{publisher}{Association for
  Computing Machinery}, \bibinfo{address}{New York, NY, USA},
  \bibinfo{pages}{231–239}.
\newblock
\showISBNx{0897912632}
\urldef\tempurl%
\url{https://doi.org/10.1145/308386.308446}
\showDOI{\tempurl}


\bibitem[\protect\citeauthoryear{K\"{o}stler, Kiessling, Th\"{o}ne, and
  G\"{u}ntzer}{K\"{o}stler et~al\mbox{.}}{1995}]%
        {datalog-subsumption}
\bibfield{author}{\bibinfo{person}{Gerhard K\"{o}stler},
  \bibinfo{person}{Werner Kiessling}, \bibinfo{person}{Helmut Th\"{o}ne}, {and}
  \bibinfo{person}{Ulrich G\"{u}ntzer}.} \bibinfo{year}{1995}\natexlab{}.
\newblock \showarticletitle{Fixpoint Iteration with Subsumption in Deductive
  Databases}.
\newblock \bibinfo{journal}{\emph{J. Intell. Inf. Syst.}} \bibinfo{volume}{4},
  \bibinfo{number}{2} (\bibinfo{date}{mar} \bibinfo{year}{1995}),
  \bibinfo{pages}{123–148}.
\newblock
\showISSN{0925-9902}
\urldef\tempurl%
\url{https://doi.org/10.1007/BF00961871}
\showDOI{\tempurl}


\bibitem[\protect\citeauthoryear{Krishnamurthy and Naqvi}{Krishnamurthy and
  Naqvi}{1988}]%
        {choice-datalog}
\bibfield{author}{\bibinfo{person}{Ravi Krishnamurthy} {and}
  \bibinfo{person}{Shamim~A. Naqvi}.} \bibinfo{year}{1988}\natexlab{}.
\newblock \showarticletitle{Non-Deterministic Choice in Datalog}. In
  \bibinfo{booktitle}{\emph{JCDKB}}.
\newblock


\bibitem[\protect\citeauthoryear{Lattner and Adve}{Lattner and Adve}{2004}]%
        {llvm}
\bibfield{author}{\bibinfo{person}{Chris Lattner} {and} \bibinfo{person}{Vikram
  Adve}.} \bibinfo{year}{2004}\natexlab{}.
\newblock \showarticletitle{LLVM: a compilation framework for lifelong program
  analysis \& transformation}. In \bibinfo{booktitle}{\emph{International
  Symposium on Code Generation and Optimization, 2004. CGO 2004.}}
  \bibinfo{pages}{75--86}.
\newblock
\urldef\tempurl%
\url{https://doi.org/10.1109/CGO.2004.1281665}
\showDOI{\tempurl}


\bibitem[\protect\citeauthoryear{Madsen, Yee, and Lhot\'{a}k}{Madsen
  et~al\mbox{.}}{2016}]%
        {flix}
\bibfield{author}{\bibinfo{person}{Magnus Madsen}, \bibinfo{person}{Ming-Ho
  Yee}, {and} \bibinfo{person}{Ond\v{r}ej Lhot\'{a}k}.}
  \bibinfo{year}{2016}\natexlab{}.
\newblock \showarticletitle{From Datalog to Flix: A Declarative Language for
  Fixed Points on Lattices}.
\newblock \bibinfo{journal}{\emph{SIGPLAN Not.}} \bibinfo{volume}{51},
  \bibinfo{number}{6} (\bibinfo{date}{jun} \bibinfo{year}{2016}),
  \bibinfo{pages}{194–208}.
\newblock
\showISSN{0362-1340}
\urldef\tempurl%
\url{https://doi.org/10.1145/2980983.2908096}
\showDOI{\tempurl}


\bibitem[\protect\citeauthoryear{Milner}{Milner}{1978}]%
        {hindley-milner}
\bibfield{author}{\bibinfo{person}{Robin Milner}.}
  \bibinfo{year}{1978}\natexlab{}.
\newblock \showarticletitle{A theory of type polymorphism in programming}.
\newblock \bibinfo{journal}{\emph{J. Comput. System Sci.}}
  \bibinfo{volume}{17}, \bibinfo{number}{3} (\bibinfo{year}{1978}),
  \bibinfo{pages}{348--375}.
\newblock
\showISSN{0022-0000}
\urldef\tempurl%
\url{https://doi.org/10.1016/0022-0000(78)90014-4}
\showDOI{\tempurl}


\bibitem[\protect\citeauthoryear{Nandi, Willsey, Anderson, Wilcox, Darulova,
  Grossman, and Tatlock}{Nandi et~al\mbox{.}}{2020}]%
        {szalinski}
\bibfield{author}{\bibinfo{person}{Chandrakana Nandi}, \bibinfo{person}{Max
  Willsey}, \bibinfo{person}{Adam Anderson}, \bibinfo{person}{James~R. Wilcox},
  \bibinfo{person}{Eva Darulova}, \bibinfo{person}{Dan Grossman}, {and}
  \bibinfo{person}{Zachary Tatlock}.} \bibinfo{year}{2020}\natexlab{}.
\newblock \showarticletitle{Synthesizing Structured {CAD} Models with Equality
  Saturation and Inverse Transformations}. In
  \bibinfo{booktitle}{\emph{Proceedings of the 41st ACM SIGPLAN Conference on
  Programming Language Design and Implementation}} (London, UK)
  \emph{(\bibinfo{series}{PLDI 2020})}. \bibinfo{publisher}{Association for
  Computing Machinery}, \bibinfo{address}{New York, NY, USA},
  \bibinfo{pages}{31–44}.
\newblock
\showISBNx{9781450376136}
\urldef\tempurl%
\url{https://doi.org/10.1145/3385412.3386012}
\showDOI{\tempurl}


\bibitem[\protect\citeauthoryear{Nandi, Willsey, Zhu, Wang, Saiki, Anderson,
  Schulz, Grossman, and Tatlock}{Nandi et~al\mbox{.}}{2021}]%
        {ruler}
\bibfield{author}{\bibinfo{person}{Chandrakana Nandi}, \bibinfo{person}{Max
  Willsey}, \bibinfo{person}{Amy Zhu}, \bibinfo{person}{Yisu~Remy Wang},
  \bibinfo{person}{Brett Saiki}, \bibinfo{person}{Adam Anderson},
  \bibinfo{person}{Adriana Schulz}, \bibinfo{person}{Dan Grossman}, {and}
  \bibinfo{person}{Zachary Tatlock}.} \bibinfo{year}{2021}\natexlab{}.
\newblock \showarticletitle{Rewrite Rule Inference Using Equality Saturation}.
\newblock \bibinfo{journal}{\emph{Proc. ACM Program. Lang.}}
  \bibinfo{volume}{5}, \bibinfo{number}{OOPSLA}, Article
  \bibinfo{articleno}{119} (\bibinfo{date}{oct} \bibinfo{year}{2021}),
  \bibinfo{numpages}{28}~pages.
\newblock
\urldef\tempurl%
\url{https://doi.org/10.1145/3485496}
\showDOI{\tempurl}


\bibitem[\protect\citeauthoryear{Nappa, Zhao, Subotić, and Scholz}{Nappa
  et~al\mbox{.}}{2019}]%
        {eqrel}
\bibfield{author}{\bibinfo{person}{Patrick Nappa}, \bibinfo{person}{David
  Zhao}, \bibinfo{person}{Pavle Subotić}, {and} \bibinfo{person}{Bernhard
  Scholz}.} \bibinfo{year}{2019}\natexlab{}.
\newblock \showarticletitle{Fast Parallel Equivalence Relations in a Datalog
  Compiler}. In \bibinfo{booktitle}{\emph{2019 28th International Conference on
  Parallel Architectures and Compilation Techniques (PACT)}}.
  \bibinfo{pages}{82--96}.
\newblock
\urldef\tempurl%
\url{https://doi.org/10.1109/PACT.2019.00015}
\showDOI{\tempurl}


\bibitem[\protect\citeauthoryear{Nelson}{Nelson}{1980}]%
        {nelson}
\bibfield{author}{\bibinfo{person}{Charles~Gregory Nelson}.}
  \bibinfo{year}{1980}\natexlab{}.
\newblock \emph{\bibinfo{title}{Techniques for Program Verification}}.
\newblock \bibinfo{thesistype}{Ph.D. Dissertation}. \bibinfo{school}{Stanford
  University}, \bibinfo{address}{Stanford, CA, USA}.
\newblock
\newblock
\shownote{AAI8011683.}


\bibitem[\protect\citeauthoryear{Ngo, Porat, R{\'e}, and Rudra}{Ngo
  et~al\mbox{.}}{2018}]%
        {generic-join}
\bibfield{author}{\bibinfo{person}{Hung~Q Ngo}, \bibinfo{person}{Ely Porat},
  \bibinfo{person}{Christopher R{\'e}}, {and} \bibinfo{person}{Atri Rudra}.}
  \bibinfo{year}{2018}\natexlab{}.
\newblock \showarticletitle{Worst-case optimal join algorithms}.
\newblock \bibinfo{journal}{\emph{Journal of the ACM (JACM)}}
  \bibinfo{volume}{65}, \bibinfo{number}{3} (\bibinfo{year}{2018}),
  \bibinfo{pages}{1--40}.
\newblock


\bibitem[\protect\citeauthoryear{Nieuwenhuis and Oliveras}{Nieuwenhuis and
  Oliveras}{2005}]%
        {pp-congr}
\bibfield{author}{\bibinfo{person}{Robert Nieuwenhuis} {and}
  \bibinfo{person}{Albert Oliveras}.} \bibinfo{year}{2005}\natexlab{}.
\newblock \showarticletitle{Proof-Producing Congruence Closure}. In
  \bibinfo{booktitle}{\emph{Proceedings of the 16th International Conference on
  Term Rewriting and Applications}} (Nara, Japan)
  \emph{(\bibinfo{series}{RTA’05})}. \bibinfo{publisher}{Springer-Verlag},
  \bibinfo{address}{Berlin, Heidelberg}, \bibinfo{pages}{453–468}.
\newblock
\showISBNx{3540255966}
\urldef\tempurl%
\url{https://doi.org/10.1007/978-3-540-32033-3\_33}
\showDOI{\tempurl}


\bibitem[\protect\citeauthoryear{Panchekha, Sanchez-Stern, Wilcox, and
  Tatlock}{Panchekha et~al\mbox{.}}{2015}]%
        {herbie}
\bibfield{author}{\bibinfo{person}{Pavel Panchekha}, \bibinfo{person}{Alex
  Sanchez-Stern}, \bibinfo{person}{James~R. Wilcox}, {and}
  \bibinfo{person}{Zachary Tatlock}.} \bibinfo{year}{2015}\natexlab{}.
\newblock \showarticletitle{Automatically Improving Accuracy for Floating Point
  Expressions}.
\newblock \bibinfo{journal}{\emph{SIGPLAN Not.}} \bibinfo{volume}{50},
  \bibinfo{number}{6} (\bibinfo{date}{June} \bibinfo{year}{2015}),
  \bibinfo{pages}{1–11}.
\newblock
\showISSN{0362-1340}
\urldef\tempurl%
\url{https://doi.org/10.1145/2813885.2737959}
\showDOI{\tempurl}


\bibitem[\protect\citeauthoryear{Polozov and Gulwani}{Polozov and
  Gulwani}{2015}]%
        {flashmeta}
\bibfield{author}{\bibinfo{person}{Oleksandr Polozov} {and}
  \bibinfo{person}{Sumit Gulwani}.} \bibinfo{year}{2015}\natexlab{}.
\newblock \showarticletitle{FlashMeta: A Framework for Inductive Program
  Synthesis}.
\newblock \bibinfo{journal}{\emph{SIGPLAN Not.}} \bibinfo{volume}{50},
  \bibinfo{number}{10} (\bibinfo{date}{oct} \bibinfo{year}{2015}),
  \bibinfo{pages}{107–126}.
\newblock
\showISSN{0362-1340}
\urldef\tempurl%
\url{https://doi.org/10.1145/2858965.2814310}
\showDOI{\tempurl}


\bibitem[\protect\citeauthoryear{Ross and Sagiv}{Ross and Sagiv}{1992}]%
        {mono-agg}
\bibfield{author}{\bibinfo{person}{Kenneth~A. Ross} {and}
  \bibinfo{person}{Yehoshua Sagiv}.} \bibinfo{year}{1992}\natexlab{}.
\newblock \showarticletitle{Monotonic Aggregation in Deductive Databases}. In
  \bibinfo{booktitle}{\emph{Proceedings of the Eleventh ACM
  SIGACT-SIGMOD-SIGART Symposium on Principles of Database Systems}} (San
  Diego, California, USA) \emph{(\bibinfo{series}{PODS '92})}.
  \bibinfo{publisher}{Association for Computing Machinery},
  \bibinfo{address}{New York, NY, USA}, \bibinfo{pages}{114–126}.
\newblock
\showISBNx{0897915194}
\urldef\tempurl%
\url{https://doi.org/10.1145/137097.137852}
\showDOI{\tempurl}


\bibitem[\protect\citeauthoryear{Rust}{Rust}{[n.d.]}]%
        {rust}
\bibfield{author}{\bibinfo{person}{Rust}.} \bibinfo{year}{[n.d.]}\natexlab{}.
\newblock \bibinfo{title}{Rust programming language}.
\newblock \bibinfo{howpublished}{\url{https://www.rust-lang.org/}}.
\newblock
\urldef\tempurl%
\url{https://www.rust-lang.org/}
\showURL{%
\tempurl}


\bibitem[\protect\citeauthoryear{Sahebolamri, Gilray, and Micinski}{Sahebolamri
  et~al\mbox{.}}{2022}]%
        {ascent}
\bibfield{author}{\bibinfo{person}{Arash Sahebolamri}, \bibinfo{person}{Thomas
  Gilray}, {and} \bibinfo{person}{Kristopher Micinski}.}
  \bibinfo{year}{2022}\natexlab{}.
\newblock \showarticletitle{Seamless deductive inference via macros}. In
  \bibinfo{booktitle}{\emph{Proceedings of the 31st ACM SIGPLAN International
  Conference on Compiler Construction}}. \bibinfo{pages}{77--88}.
\newblock


\bibitem[\protect\citeauthoryear{Schleich, Shaikhha, and Suciu}{Schleich
  et~al\mbox{.}}{2022}]%
        {storel}
\bibfield{author}{\bibinfo{person}{Maximilian Schleich}, \bibinfo{person}{Amir
  Shaikhha}, {and} \bibinfo{person}{Dan Suciu}.}
  \bibinfo{year}{2022}\natexlab{}.
\newblock \bibinfo{title}{Optimizing Tensor Programs on Flexible Storage}.
\newblock
\newblock
\urldef\tempurl%
\url{https://doi.org/10.48550/ARXIV.2210.06267}
\showDOI{\tempurl}


\bibitem[\protect\citeauthoryear{Smaragdakis and Bravenboer}{Smaragdakis and
  Bravenboer}{2010}]%
        {doop-datalog}
\bibfield{author}{\bibinfo{person}{Yannis Smaragdakis} {and}
  \bibinfo{person}{Martin Bravenboer}.} \bibinfo{year}{2010}\natexlab{}.
\newblock \showarticletitle{Using Datalog for Fast and Easy Program Analysis}.
  In \bibinfo{booktitle}{\emph{Proceedings of the First International
  Conference on Datalog Reloaded}} (Oxford, UK)
  \emph{(\bibinfo{series}{Datalog'10})}. \bibinfo{publisher}{Springer-Verlag},
  \bibinfo{address}{Berlin, Heidelberg}, \bibinfo{pages}{245–251}.
\newblock
\showISBNx{9783642242052}
\urldef\tempurl%
\url{https://doi.org/10.1007/978-3-642-24206-9_14}
\showDOI{\tempurl}


\bibitem[\protect\citeauthoryear{Steensgaard}{Steensgaard}{1996}]%
        {DBLP:conf/popl/Steensgaard96}
\bibfield{author}{\bibinfo{person}{Bjarne Steensgaard}.}
  \bibinfo{year}{1996}\natexlab{}.
\newblock \showarticletitle{Points-to Analysis in Almost Linear Time}. In
  \bibinfo{booktitle}{\emph{Conference Record of POPL'96: The 23rd {ACM}
  {SIGPLAN-SIGACT} Symposium on Principles of Programming Languages, Papers
  Presented at the Symposium, St. Petersburg Beach, Florida, USA, January
  21-24, 1996}}, \bibfield{editor}{\bibinfo{person}{Hans{-}Juergen Boehm} {and}
  \bibinfo{person}{Guy L.~Steele Jr.}} (Eds.). \bibinfo{publisher}{{ACM}
  Press}, \bibinfo{pages}{32--41}.
\newblock
\urldef\tempurl%
\url{https://doi.org/10.1145/237721.237727}
\showDOI{\tempurl}


\bibitem[\protect\citeauthoryear{Szab\'{o}, Bergmann, Erdweg, and
  Voelter}{Szab\'{o} et~al\mbox{.}}{2018}]%
        {inca}
\bibfield{author}{\bibinfo{person}{Tam\'{a}s Szab\'{o}},
  \bibinfo{person}{G\'{a}bor Bergmann}, \bibinfo{person}{Sebastian Erdweg},
  {and} \bibinfo{person}{Markus Voelter}.} \bibinfo{year}{2018}\natexlab{}.
\newblock \showarticletitle{Incrementalizing Lattice-Based Program Analyses in
  Datalog}.
\newblock \bibinfo{journal}{\emph{Proc. ACM Program. Lang.}}
  \bibinfo{volume}{2}, \bibinfo{number}{OOPSLA}, Article
  \bibinfo{articleno}{139} (\bibinfo{date}{oct} \bibinfo{year}{2018}),
  \bibinfo{numpages}{29}~pages.
\newblock
\urldef\tempurl%
\url{https://doi.org/10.1145/3276509}
\showDOI{\tempurl}


\bibitem[\protect\citeauthoryear{Tarjan}{Tarjan}{1975}]%
        {unionfind}
\bibfield{author}{\bibinfo{person}{Robert~Endre Tarjan}.}
  \bibinfo{year}{1975}\natexlab{}.
\newblock \showarticletitle{Efficiency of a Good But Not Linear Set Union
  Algorithm}.
\newblock \bibinfo{journal}{\emph{J. ACM}} \bibinfo{volume}{22},
  \bibinfo{number}{2} (\bibinfo{date}{April} \bibinfo{year}{1975}),
  \bibinfo{pages}{215–225}.
\newblock
\showISSN{0004-5411}
\urldef\tempurl%
\url{https://doi.org/10.1145/321879.321884}
\showDOI{\tempurl}


\bibitem[\protect\citeauthoryear{Tate, Stepp, Tatlock, and Lerner}{Tate
  et~al\mbox{.}}{2009}]%
        {eqsat}
\bibfield{author}{\bibinfo{person}{Ross Tate}, \bibinfo{person}{Michael Stepp},
  \bibinfo{person}{Zachary Tatlock}, {and} \bibinfo{person}{Sorin Lerner}.}
  \bibinfo{year}{2009}\natexlab{}.
\newblock \showarticletitle{Equality Saturation: A New Approach to
  Optimization}. In \bibinfo{booktitle}{\emph{Proceedings of the 36th Annual
  ACM SIGPLAN-SIGACT Symposium on Principles of Programming Languages}}
  (Savannah, GA, USA) \emph{(\bibinfo{series}{POPL '09})}.
  \bibinfo{publisher}{ACM}, \bibinfo{address}{New York, NY, USA},
  \bibinfo{pages}{264--276}.
\newblock
\showISBNx{978-1-60558-379-2}
\urldef\tempurl%
\url{https://doi.org/10.1145/1480881.1480915}
\showDOI{\tempurl}


\bibitem[\protect\citeauthoryear{Van~Gelder}{Van~Gelder}{1992}]%
        {agg-semantics}
\bibfield{author}{\bibinfo{person}{Allen Van~Gelder}.}
  \bibinfo{year}{1992}\natexlab{}.
\newblock \showarticletitle{The Well-Founded Semantics of Aggregation}. In
  \bibinfo{booktitle}{\emph{Proceedings of the Eleventh ACM
  SIGACT-SIGMOD-SIGART Symposium on Principles of Database Systems}} (San
  Diego, California, USA) \emph{(\bibinfo{series}{PODS '92})}.
  \bibinfo{publisher}{Association for Computing Machinery},
  \bibinfo{address}{New York, NY, USA}, \bibinfo{pages}{127–138}.
\newblock
\showISBNx{0897915194}
\urldef\tempurl%
\url{https://doi.org/10.1145/137097.137854}
\showDOI{\tempurl}


\bibitem[\protect\citeauthoryear{VanHattum, Nigam, Lee, Bornholt, and
  Sampson}{VanHattum et~al\mbox{.}}{2021}]%
        {diospyros}
\bibfield{author}{\bibinfo{person}{Alexa VanHattum}, \bibinfo{person}{Rachit
  Nigam}, \bibinfo{person}{Vincent~T. Lee}, \bibinfo{person}{James Bornholt},
  {and} \bibinfo{person}{Adrian Sampson}.} \bibinfo{year}{2021}\natexlab{}.
\newblock \bibinfo{booktitle}{\emph{Vectorization for Digital Signal Processors
  via Equality Saturation}}.
\newblock \bibinfo{publisher}{Association for Computing Machinery},
  \bibinfo{address}{New York, NY, USA}, \bibinfo{pages}{874–886}.
\newblock
\showISBNx{9781450383172}
\urldef\tempurl%
\url{https://doi.org/10.1145/3445814.3446707}
\showURL{%
\tempurl}


\bibitem[\protect\citeauthoryear{Wang, Dillig, and Singh}{Wang
  et~al\mbox{.}}{2017a}]%
        {blaze}
\bibfield{author}{\bibinfo{person}{Xinyu Wang}, \bibinfo{person}{Isil Dillig},
  {and} \bibinfo{person}{Rishabh Singh}.} \bibinfo{year}{2017}\natexlab{a}.
\newblock \showarticletitle{Program Synthesis Using Abstraction Refinement}.
\newblock \bibinfo{journal}{\emph{Proc. ACM Program. Lang.}}
  \bibinfo{volume}{2}, \bibinfo{number}{POPL}, Article \bibinfo{articleno}{63}
  (\bibinfo{date}{dec} \bibinfo{year}{2017}), \bibinfo{numpages}{30}~pages.
\newblock
\urldef\tempurl%
\url{https://doi.org/10.1145/3158151}
\showDOI{\tempurl}


\bibitem[\protect\citeauthoryear{Wang, Dillig, and Singh}{Wang
  et~al\mbox{.}}{2017b}]%
        {dace}
\bibfield{author}{\bibinfo{person}{Xinyu Wang}, \bibinfo{person}{Isil Dillig},
  {and} \bibinfo{person}{Rishabh Singh}.} \bibinfo{year}{2017}\natexlab{b}.
\newblock \showarticletitle{Synthesis of Data Completion Scripts Using Finite
  Tree Automata}.
\newblock \bibinfo{journal}{\emph{Proc. ACM Program. Lang.}}
  \bibinfo{volume}{1}, \bibinfo{number}{OOPSLA}, Article
  \bibinfo{articleno}{62} (\bibinfo{date}{oct} \bibinfo{year}{2017}),
  \bibinfo{numpages}{26}~pages.
\newblock
\urldef\tempurl%
\url{https://doi.org/10.1145/3133886}
\showDOI{\tempurl}


\bibitem[\protect\citeauthoryear{Wang, Hutchison, Leang, Howe, and Suciu}{Wang
  et~al\mbox{.}}{2020}]%
        {spores}
\bibfield{author}{\bibinfo{person}{Yisu~Remy Wang}, \bibinfo{person}{Shana
  Hutchison}, \bibinfo{person}{Jonathan Leang}, \bibinfo{person}{Bill Howe},
  {and} \bibinfo{person}{Dan Suciu}.} \bibinfo{year}{2020}\natexlab{}.
\newblock \showarticletitle{{SPORES}: Sum-Product Optimization via Relational
  Equality Saturation for Large Scale Linear Algebra}.
\newblock \bibinfo{journal}{\emph{Proceedings of the VLDB Endowment}}
  (\bibinfo{year}{2020}).
\newblock


\bibitem[\protect\citeauthoryear{Whaley, Avots, Carbin, and Lam}{Whaley
  et~al\mbox{.}}{2005}]%
        {bddbddb}
\bibfield{author}{\bibinfo{person}{John Whaley}, \bibinfo{person}{Dzintars
  Avots}, \bibinfo{person}{Michael Carbin}, {and} \bibinfo{person}{Monica~S.
  Lam}.} \bibinfo{year}{2005}\natexlab{}.
\newblock \showarticletitle{Using Datalog with Binary Decision Diagrams for
  Program Analysis}. In \bibinfo{booktitle}{\emph{Proceedings of the Third
  Asian Conference on Programming Languages and Systems}} (Tsukuba, Japan)
  \emph{(\bibinfo{series}{APLAS'05})}. \bibinfo{publisher}{Springer-Verlag},
  \bibinfo{address}{Berlin, Heidelberg}, \bibinfo{pages}{97–118}.
\newblock
\showISBNx{3540297359}
\urldef\tempurl%
\url{https://doi.org/10.1007/11575467_8}
\showDOI{\tempurl}


\bibitem[\protect\citeauthoryear{Willsey, Nandi, Wang, Flatt, Tatlock, and
  Panchekha}{Willsey et~al\mbox{.}}{2021}]%
        {egg}
\bibfield{author}{\bibinfo{person}{Max Willsey}, \bibinfo{person}{Chandrakana
  Nandi}, \bibinfo{person}{Yisu~Remy Wang}, \bibinfo{person}{Oliver Flatt},
  \bibinfo{person}{Zachary Tatlock}, {and} \bibinfo{person}{Pavel Panchekha}.}
  \bibinfo{year}{2021}\natexlab{}.
\newblock \showarticletitle{Egg: Fast and Extensible Equality Saturation}.
\newblock \bibinfo{journal}{\emph{Proc. ACM Program. Lang.}}
  \bibinfo{volume}{5}, \bibinfo{number}{POPL}, Article \bibinfo{articleno}{23}
  (\bibinfo{date}{jan} \bibinfo{year}{2021}), \bibinfo{numpages}{29}~pages.
\newblock
\urldef\tempurl%
\url{https://doi.org/10.1145/3434304}
\showDOI{\tempurl}


\bibitem[\protect\citeauthoryear{Wolfman, Domingos, and Weld}{Wolfman
  et~al\mbox{.}}{2001}]%
        {vsa}
\bibfield{author}{\bibinfo{person}{Steven Wolfman}, \bibinfo{person}{Pedro
  Domingos}, {and} \bibinfo{person}{Daniel Weld}.}
  \bibinfo{year}{2001}\natexlab{}.
\newblock \showarticletitle{Programming By Demonstration Using Version Space
  Algebra}.
\newblock \bibinfo{journal}{\emph{Machine Learning}}  \bibinfo{volume}{53}
  (\bibinfo{date}{12} \bibinfo{year}{2001}).
\newblock
\urldef\tempurl%
\url{https://doi.org/10.1023/A:1025671410623}
\showDOI{\tempurl}


\bibitem[\protect\citeauthoryear{Yang, Phothilimtha, Wang, Willsey, Roy, and
  Pienaar}{Yang et~al\mbox{.}}{2021}]%
        {tensat}
\bibfield{author}{\bibinfo{person}{Yichen Yang},
  \bibinfo{person}{Phitchaya~Mangpo Phothilimtha}, \bibinfo{person}{Yisu~Remy
  Wang}, \bibinfo{person}{Max Willsey}, \bibinfo{person}{Sudip Roy}, {and}
  \bibinfo{person}{Jacques Pienaar}.} \bibinfo{year}{2021}\natexlab{}.
\newblock \showarticletitle{Equality Saturation for Tensor Graph
  Superoptimization}. In \bibinfo{booktitle}{\emph{Proceedings of Machine
  Learning and Systems}}.
\newblock
\showeprint{2101.01332}


\bibitem[\protect\citeauthoryear{Zhang, Wang, Flatt, Cao, Zucker, Rosenthal,
  Tatlock, and Willsey}{Zhang et~al\mbox{.}}{2023}]%
        {egglog-preprint}
\bibfield{author}{\bibinfo{person}{Yihong Zhang}, \bibinfo{person}{Yisu~Remy
  Wang}, \bibinfo{person}{Oliver Flatt}, \bibinfo{person}{David Cao},
  \bibinfo{person}{Philip Zucker}, \bibinfo{person}{Eli Rosenthal},
  \bibinfo{person}{Zachary Tatlock}, {and} \bibinfo{person}{Max Willsey}.}
  \bibinfo{year}{2023}\natexlab{}.
\newblock \bibinfo{title}{Better Together: Unifying Datalog and Equality
  Saturation}.
\newblock
\newblock
\showeprint[arxiv]{2304.04332}~[cs.PL]


\bibitem[\protect\citeauthoryear{Zhang, Wang, Willsey, and Tatlock}{Zhang
  et~al\mbox{.}}{2022}]%
        {relational-ematching}
\bibfield{author}{\bibinfo{person}{Yihong Zhang}, \bibinfo{person}{Yisu~Remy
  Wang}, \bibinfo{person}{Max Willsey}, {and} \bibinfo{person}{Zachary
  Tatlock}.} \bibinfo{year}{2022}\natexlab{}.
\newblock \showarticletitle{Relational E-Matching}.
\newblock \bibinfo{journal}{\emph{Proc. ACM Program. Lang.}}
  \bibinfo{volume}{6}, \bibinfo{number}{POPL}, Article \bibinfo{articleno}{35}
  (\bibinfo{date}{jan} \bibinfo{year}{2022}), \bibinfo{numpages}{22}~pages.
\newblock
\urldef\tempurl%
\url{https://doi.org/10.1145/3498696}
\showDOI{\tempurl}


\end{thebibliography}
